\documentclass[%
twocolumn,
groupedaddress,
amsmath,amssymb,
aps,
pra,
]{revtex4-2}

\usepackage{graphicx}
\usepackage{amsfonts}
\usepackage{bm}
\usepackage{hyperref,url}

\usepackage[dvipsnames]{xcolor}
\usepackage{soul}

\newcommand{\bd}{\begin{displaymath}}
\newcommand{\ed}{\end{displaymath}}
\newcommand{\be}{\begin{equation}}
\newcommand{\ee}{\end{equation}}
\newcommand{\bs}{\begin{subequations}}
\newcommand{\es}{\end{subequations}}
\newcommand{\ba}{\begin{eqnarray}}
\newcommand{\ea}{\end{eqnarray}}

\begin{document}

\title{Bohmian analysis of dark solutions in interfering Bose-Einstein condensates:\\
the dynamical role of underlying velocity fields}

\author{J. Tounli}
\affiliation{Department of Optics, Faculty of Physical Sciences,
Universidad Complutense de Madrid\\
Pza.\ Ciencias 1, Ciudad Universitaria -- 28040 Madrid, Spain}

\author{A. S. Sanz}
\email{Corresponding author: a.s.sanz@fis.ucm.es}
\affiliation{Department of Optics, Faculty of Physical Sciences,
Universidad Complutense de Madrid\\
Pza.\ Ciencias 1, Ciudad Universitaria -- 28040 Madrid, Spain}

\date{\today}

\begin{abstract}
In the last decades, the experimental research on Bose-Einstein interferometry has
received much attention due to promising technological implications.
This has thus motivated the development of numerical simulations aimed at solving the
time-dependent Gross-Pitaevskii equation and its reduced one-dimensional version to better
understand the development of interference-type features and the subsequent soliton dynamics.
In this work, Bohmian mechanics is considered as an additional tool to further explore and
analyze the formation and evolution in real time of the soliton arrays that follow the
merging of two condensates.
An alternative explanation is thus provided in terms of an underlying dynamical velocity
field, directly linked to the local phase variations undergone by the condensate along
its evolution.
Although the reduced one-dimensional model is considered here, it still captures the essence
of the phenomenon, rendering a neat picture of the full evolution without diminishing the
generality of the description.
To better appreciate the subtleties of free versus bound dynamics, two cases are discussed.
First, the soliton dynamics exhibited by a coherent superposition of two freely released
condensates is studied, discussing the peculiarities of the underlying velocity field and
the corresponding flux trajectories in terms of both the peak-to-peak distance between
the two initial clouds and the addition of a phase difference between them.
In the latter case, an interesting correspondence with the well-known Aharonov-Bohm effect
is found.
Then, the recurrence dynamics displayed by the more general case of two condensates released
from the two opposite turning points of a harmonic trap is considered in terms of the
distance between such turning points.
In both cases, it is presumed that the initial superposition state is generated by splitting
adiabatically a single condensate with the aid of an optical lattice, which is then turned
off.
Nonetheless, although the lattice does not play any active role in the simulations, the
parameters defining the initial states are in compliance with it, which helps in the
interpretation and understanding of the results observed.
\end{abstract}

\maketitle


\section{Introduction}
\label{sec1}
\vspace{-.25cm}

In the late 1990s, Ketterle and coworkers \cite{Andrews637,ketterle}
produced the first experimental realization of interference with a
Bose-Einstein condensate (BEC).
As they showed, when an ultracold atomic cloud is coherently split up and
then the two resulting separate clouds are released again, the latter interact
in such a way that a pattern of alternating bands of more and less atomic
density can be observed.
To some extent, an analogy can be established between this behavior and the
interference fringes observed in a typical Young-type two-slit experiment,
although the latter case obeys a linear dynamics, while the former is a
consequence of a nonlinear one.
A similar effect can also be observed by splitting the condensate and then
letting one of the parties to drop on top of the other, as it was shown by
Javanainen \textit{et al.}~\cite{javanainen1997phase,javanainen1996quantum}
by the same time.
Since the performance of these crucial experiments, interferometry with BECs
\cite{dalfovo1999theory,pitaevskii2016bose,pethick2008bose, PhysRevLett.100.080405} has
become an important test ground for the understanding of quantum coherence
as well as a remarkable source of novel quantum technology, because of its
extraordinary sensitivity, with applications in atom lasers \cite{mewes} or
quantum metrology \cite{Lee2012}, where the manipulation and
preparation of BECs in optical lattices plays a major role
\cite{PhysRevA.66.021601,Morsch2002}.

In order to understand and describe the dynamics exhibited by BECs in real time,
the Gross-Pitaevskii equation (GPE) constitutes an ideal and simple working tool,
without requiring a further many-body description of the full atomic cloud
(involving of the order of $10^3$ to $10^6$ atoms, in general terms).
This is possible by recasting the many-body problem in the form of an effective
nonlinear Schr\"odiner equation, which accounts for the dynamics displayed by a single
atom from the cloud acted by an external potential plus a self-interaction term that
represents the collective action of any other identical atom from the cloud
\cite{dalfovo1999theory,pitaevskii2016bose,pethick2008bose}.
Because the GPE is not analytical in general, a number of numerical
techniques have been considered in the literature to solve this equation and extract
useful information from it \cite{bao2003numerical,ANTOINE20132621,PhysRevE.62.1382}.
To some extent, these techniques are analogous to those earlier on applied to solve
the time-dependent Schr\"odinger equation, and have been used recently to study,
analyze and describe different aspects involved in BEC interferometry
\cite{polo2013soliton,Ji_2016,ahufinger1,ahufinger2}, the BEC dynamics in periodic and
harmonic potentials \cite{PhysRevE.75.036214}, or the role of the nonlinearity when
weakly harmonic and Gaussian traps are considered
\cite{PhysRevE.85.056608,PhysRevE.89.013204}.

Within this context, a suitable tool to understand the dynamics displayed in real time
by BECs is the Bohmian formulation of quantum mechanics
\cite{sanz-bk-1,sanz:FrontPhys:2019}.
The hydrodynamic language introduced by this quantum formulation allows
us to understand the interference dynamics leading to the appearance of
solitons in terms of an underlying velocity field, associated with the
phase of the condensate, and to follow its subsequent evolution in time
by means of swarms of trajectories.
It was shown by Benseny {\it et al.}~\cite{benseny:PanStanford:2012} that
these trajectories constitute a convenient tool to determine how each
element of the condensate (not to be confused with each individual atom itself)
moves apart, thus providing some clues on its dynamical
evolution beyond more conventional-type information, such as the
density distribution.
In other words, it is possible to determine which portions in the BEC
are going to separate faster or slower at each time by simply observing
local variations in the underlying velocity field.

In this work we investigate by means of a series of numerical simulations
the formation of dark solitons both from two freely released condensates and also under
the action of an underlying harmonic trap in order to study the appearance and
persistence of recurrences in time.
The appearance of soliton-type solutions is associated with the value of the
scattering length, $a_s$.
Specifically, if $a_s < 0$, as it happens in $^{85}$Rb \cite{PhysRevLett.85.1795} or $^7$Li \cite{PhysRevLett.75.1687}, atomic interactions are attractive and bright soliton solutions (spike-type deformations in the density) can appear.
On the contrary, for $a_s > 0$, as it is the case of $^{87}$Rb \cite{Anderson198}
or $^{23}$Na \cite{PhysRevLett.75.3969}, the interactions are repulsive, which
corresponds to a situation where dark solitons (dips in the density) can be observed.
The first experimental observation of dark solitons dates back to 25 years ago, where these solitons were produced in an elongated (cigar shaped) BEC by the so-called phase imprinting method \cite{lewenstein:PRL:1999}.
A year before, though, Scott {\it et al.}~\cite{burnett:JPB:1998} identified the formation of persistent dark fringes in the the case of the collision of two separated condensates
under the influence of a harmonic trap.
This work showed evidence of the relationship between the dark solitons and the phase change, an idea that is implicit in the phase imprinting method devised to generate vortices in the condentasate \cite{dobrek:PRA:1999}, but that also led to the experimental generation of dark solitons \cite{lewenstein:PRL:1999}.
An overview of the advances in dark soliton dynamics both experimentally and theoretically can be found in \cite{Frantzeskakis-bk:2008,Frantzeskakis:JPA:2010}.
Such phase imprinting methods are actually strongly connected to the Bohmian trajectories here.
As it is shown below, these trajectories will allow us to monitor in real time some features that are not evident at a first glance from the usual density distributions, and also to elucidate some important differences between the nonlinear and linear dynamical regimes that
characterize the BEC dynamics in harmonic traps.
This is by virtue of the relationship between the trajectories and the local phase of the condensate.
Indeed, pioneering work by Tsuzuki \cite{tsuzuki:JLTP:1971} in the early 1970s on the derivation of soliton-like solutions of the Gross-Pitaevskii equation are strongly connected to this formulation (Tsuzuki's is only a first approximation to the exact approach here considered).

The work has been organized as follows.
In Sec.~\ref{sec2} the model and computational details are introduced as well
as some elementary notions of the Bohmian formulation.
In Sec.~\ref{sec3} the results obtained from the numerical simulations carried
out for the two scenarios mentioned above, namely, free propagation and motion
inside a harmonic trap, are presented and discussed.
To conclude, some final remarks are summarized in Sec.~\ref{sec4}.


\vspace{-.25cm}
\section{Theory}
\label{sec2}
\vspace{-.25cm}


\subsection{The reduced one-dimensional GPE}
\label{sec21}
\vspace{-.25cm}

Within the Hartree-Fock approximation, the dynamics of a BEC consisting of $N$ identical atoms
with mass $m$ is described by the Gross-Pitaevskii equation \cite{dalfovo1999theory},
\be
 i \hbar \frac{\partial \Psi ({\bf r}, t)}{\partial t} =  \left[ - \frac{\hbar^2}{2m}\ \nabla^2
  + V_{\rm ext} ({\bf r},t) + g \mathcal{N}({\bf r},t) \right] \Psi ({\bf r}, t) ,
 \label{eq:1.2}
\ee
where $V_{\rm ext}$ describes any external interaction acting on the BEC (e.g., the
confining optical trap), while the nonlinear term $g\mathcal{N}({\bf r},t)$ accounts
for the self-interaction contribution, i.e., the effective interaction with the
remaining $N-1$ atoms in the cloud.
The strength of such interaction is determined by the coupling constant
$g = 4\pi\hbar^2 a_s/m$, with $a_s$ being the scattering length \cite{pethick2008bose}.
For convenience, the wave function (order parameter) can be recast in polar form, as
\be
 \Psi({\bf r}, t) = \sqrt{\mathcal{N}({\bf r}, t)} e^{i \theta({\bf r}, t)} ,
 \label{eq2}
\ee
which enables a description of the condensate in terms of its density distribution,
$\mathcal{N}({\bf r},t) = |\Psi ({\bf r}, t)|^2$, with
$\int |\Psi ({\bf r}, t)|^2 d{\bf r} = N$, and its local phase variations, accounted
for by $\theta({\bf r}, t)$.
Here, for numerical convenience, we have chosen the wave function to be normalized
to unity, which implies that the factor $N$ arising from $\mathcal{N}({\bf r}, t)$
will be included as a multiplicative constant in $g$, that is, from now on 
$g = 4\pi\hbar^2 a_s N/m$.

Consider that a prolate configuration for the condensate, which arises assuming typical
frequency ranges $f_z \sim 20-60$~Hz and $f_\perp \sim 400-900$~Hz.
Accordingly, the transverse harmonic trap values will be much larger than the longitudinal
one, since $\omega_\perp^2 \gg \omega_z^2$, and hence, for the times considered, the
transverse degrees of freedom can be assumed to be frozen.
This thus allows us to provide an effective description of the condensate dynamics along
the $z$-direction by means of a reduced one-dimensional (1D) GPE \cite{pethick2008bose},
which reads as
\be
 i\hbar \frac{\partial \psi (z,t)}{\partial t} = \left[ -\frac{\hbar^2}{2m}\frac{\partial^2}{\partial z^2}
  + V_{\rm ext} (z) + g_{1D} n(z,t) \right] \psi(z,t) ,
 \label{GPE1D}
\ee
where $g_{1D} = g/2\pi a_\perp^2 = 2\hbar \omega_\perp a_s$, with
$a_\perp = \sqrt{\hbar/m\omega_\perp}$, is an effective 1D coupling constant that arises
from the dimensional reduction of Eq.~(\ref{eq:1.2}) \cite{parker2003soliton,ronzheimer:PhDThesis:2008} (see
further details about the potential parameters in Sec.~\ref{sec21b}).
From now on, for computational convenience we assume that the wave function $\psi(z,t)$ is
normalized to unity, i.e., $\int |\psi (z,t)|^2 dz = \int n(z,t) dz = 1$, with $n(z,t)$ being
a linear density distribution (measured in $\mu$m$^{-1}$).
Accordingly, the expression for the coupling constant $g_{1D}$ here will read as
$g_{1D} = g'/4\pi a_\perp^2 = 2 \hbar \omega_\perp a_s N$, with $a_\perp = \sqrt{\hbar/m\omega_\perp}$
(to further simplify notation, $g_{1D}$ is used instead of $g'_{1D}$).

Typical single dark soliton solutions for this nonlinear equation display the functional form
\cite{parker2003soliton}
\be
 \psi(z, t) = \sqrt{n_0} e^{-i (\mu / \hbar) t}
  \left[ \beta \tanh \left[ \beta \frac{(z - vt)}{\chi} \right] + i\frac{v}{c} \right]
 \label{solit1}
\ee
for BEC clouds with a constant density $n_0$, where $\mu = n_0 g $ is the chemical potential,
$c = \sqrt{\mu/m}$ is the so-called Bogoliubov speed of sound, $\beta = \sqrt{1 - (v/c)^2}$
and $\chi = \hbar/\sqrt{m n_0 g}$ is the coherence or healing length characterizing the soliton
extension.
The solution (\ref{solit1}) represents a soliton propagating towards positive
$z$ with a speed $v$ on a homogeneous background density $n_0$.
As it can be readily inferred from Eq.~(\ref{solit1}), in the particular case $v = c$, the
soliton will not be distinguishable from the background fluid \cite{parker2010dark}.
Here, instead of a constant density, we analyze the case of space-limited condensates with
an initial Gaussian profile, which somehow mimic the typical parabolic density profile that
corresponds to a stationary solution inside a harmonic trap~\cite{pethick2008bose}.
Note that solutions like (\ref{solit1}) can be generated from the Gaussian ansatz by
imprinting a phase shift to a part of the condensate leaving the other part unaffected
\cite{ahufinger1}.
Following this basic imprinting technique, the appearance of interference-type features
in BECs can be understood as a sequential phase change, which allows us to identify the deeps
in the cloud with a sequence of dark solitions, each with a shape analogous to
Eq.~(\ref{solit1}).
This will be made evident by means of the Bohmian analysis.


\vspace{-.25cm}
\subsection{Bohmian description}
\label{sec22}
\vspace{-.25cm}

A central idea to hydrodynamics is that fluid diffusion can be monitored by means of
streamlines, which provide us with information on the expansion, contraction, or rotation
of the fluid.
In a similar fashion, the Bohmian formulation of quantum mechanics or Bohmian mechanics
renders a trajectory-based description of the evolution of quantum systems or, more
strictly speaking, the diffusion of their probability density in the corresponding
configuration space \cite{sanz:FrontPhys:2019}.
In its original version, either as formulated by Madelung in 1926 \cite{madelung:ZPhys:1926}
or, later on, as postulated by Bohm in 1952 \cite{bohm:PR:1952-1,bohm:PR:1952-2}, the main equations of motion are obtained after recasting the wave function in polar form (i.e.,
considering a nonlinear transformation from a complex-valued field to two real-valued fields)
and then substituting it into the time-dependent Schr\"odinger equation.
Proceeding the same way, i.e., substituting the polar ansatz (\ref{eq2}) into 
Eq.~(\ref{eq:1.2}), leads to the set of coupled hydrodynamic equations of motion
\bs
\ba
 \frac{\partial \mathcal{N}({\bf r},t)}{\partial t} & = &
  - \nabla \cdot {\bf j} ({\bf r},t) ,
 \label{conserv} \\
 \hbar\ \frac{\partial \theta ({\bf r},t)}{\partial t} & + & \frac{\hbar^2}{2m} \left( \nabla \theta \right)^2 + V_{\rm ext} ({\bf r}) \nonumber \\
  & & + g \mathcal{N}({\bf r},t) + Q ({\bf r},t) = 0 ,
 \label{HJE}
\ea
 \label{bohmeqs}
\es
where
\ba
 {\bf j} ({\bf r},t) & \equiv &
 \frac{\hbar}{m}\ \mathcal{N}({\bf r},t) \nabla \theta
 \nonumber \\
 & = & \frac{\hbar}{2mi} \left[ \Psi^* ({\bf r},t) \nabla \Psi ({\bf r},t)
  - \Psi ({\bf r},t) \nabla \Psi^* ({\bf r},t) \right] \nonumber \\
\ea
is the quantum flux \cite{schiff-bk} associated with the process, which ensures the conservation of particles in the cloud by virtue of Eq.~(\ref{conserv}).
Equation~(\ref{HJE}) displays the functional form of a Hamilton-Jacobi equation, although with
the particularity that it contains the term
\be
 Q ({\bf r},t) = - \frac{\hbar^2}{4m}
 \left\{ \frac{\nabla^2 \mathcal{N}({\bf r},t)}{\mathcal{N}({\bf r},t)} - \frac{1}{2} \left[ \frac{\nabla \mathcal{N}({\bf r},t)}{\mathcal{N}({\bf r},t)} \right]^2 \right\} ,
 \label{QP}
\ee
which is the so-called Bohm's quantum potential \cite{holland-bk}.
As it can be noticed, unlike the motion described by a standard Hamilton-Jacobi equation, the
presence of this additional term is going to induce a permanent coupling between the motion
displayed by a single particle of mass $m$ and a statistical ensemble of identical particles,
specified by the density $n$.

From Eqs.~(\ref{bohmeqs}) it is clear that individual trajectories or streamlines associated
with the evolution of the quantum system described by (\ref{eq:1.2}) can be obtained either
by postulating an equation of motion from (\ref{HJE}) \cite{bohm:PR:1952-1,holland-bk},
in direct analogy with the classical counterpart,
\be
 \dot{\bf r} ({\bf r},t) = \frac{\hbar}{m}\ \nabla \theta ({\bf r},t) ,
 \label{mot1}
\ee
or just, in a more natural way, by noting from the continuity equation (\ref{conserv}) that the
quantum flux is typically associated with a local velocity field \cite{sanz:FrontPhys:2019},
\be
 {\bf j} ({\bf r},t) = \mathcal{N}({\bf r},t)  {\bf v} ({\bf r},t) ,
\ee
from which the same equation of motion follows,
\be
 {\bf v} ({\bf r},t) = \dot{\bf r} ({\bf r},t) =
 \frac{{\bf j} ({\bf r},t)}{\mathcal{N}({\bf r},t)}
  = \frac{1}{m}\ {\rm Re} \left\{ \frac{\hat{\bf p} \Psi ({\bf r},t)}
   {\Psi ({\bf r},t)} \right\} .
 \label{mot2}
\ee
The last term, emphasizes the direct connection of this velocity vector field with the real
part of the local value of the usual momentum operator $\hat{\bf p} = - i\hbar \nabla$.
Hence, the trajectories obtained in this manner, i.e., from Eq.~(\ref{mot2}), are well-defined and are not in contradiction with standard quantum mechanics, where, in principle, one cannot appeal to the concept of trajectory in the configuration space, because of the uncertainty relation between position and momentum.
Unlike classical trajectories, the swarms of trajectories obtained from Eq.~(\ref{mot2}) are constrained to obey a phase relation, according to which the corresponding momenta cannot be independent one another.
This phase relation is precisely what we regard as coherence, which, in the present context, gives rise to the well-known Bohmian trajectory non-crossing rule: Bohmian trajectories cannot get across the same point at the same time, this being a direct consequence from the single-valuedness of the local phase of quantum systems.
Besides, it is worth stressing the fact that the equation of motion (\ref{mot2}) not only establishes a direct link with usual transport equations, but also with the hydrodynamical description of Bose fluids earlier on proposed by Landau \cite{landau:JPUSSR:1941,landau:PhysRev:1941}.


\vspace{-.25cm}
\subsection{Trapping potential model}
\label{sec21b}
\vspace{-.25cm}

\begin{figure}[!t]
 \centering
 \includegraphics[width=\columnwidth]{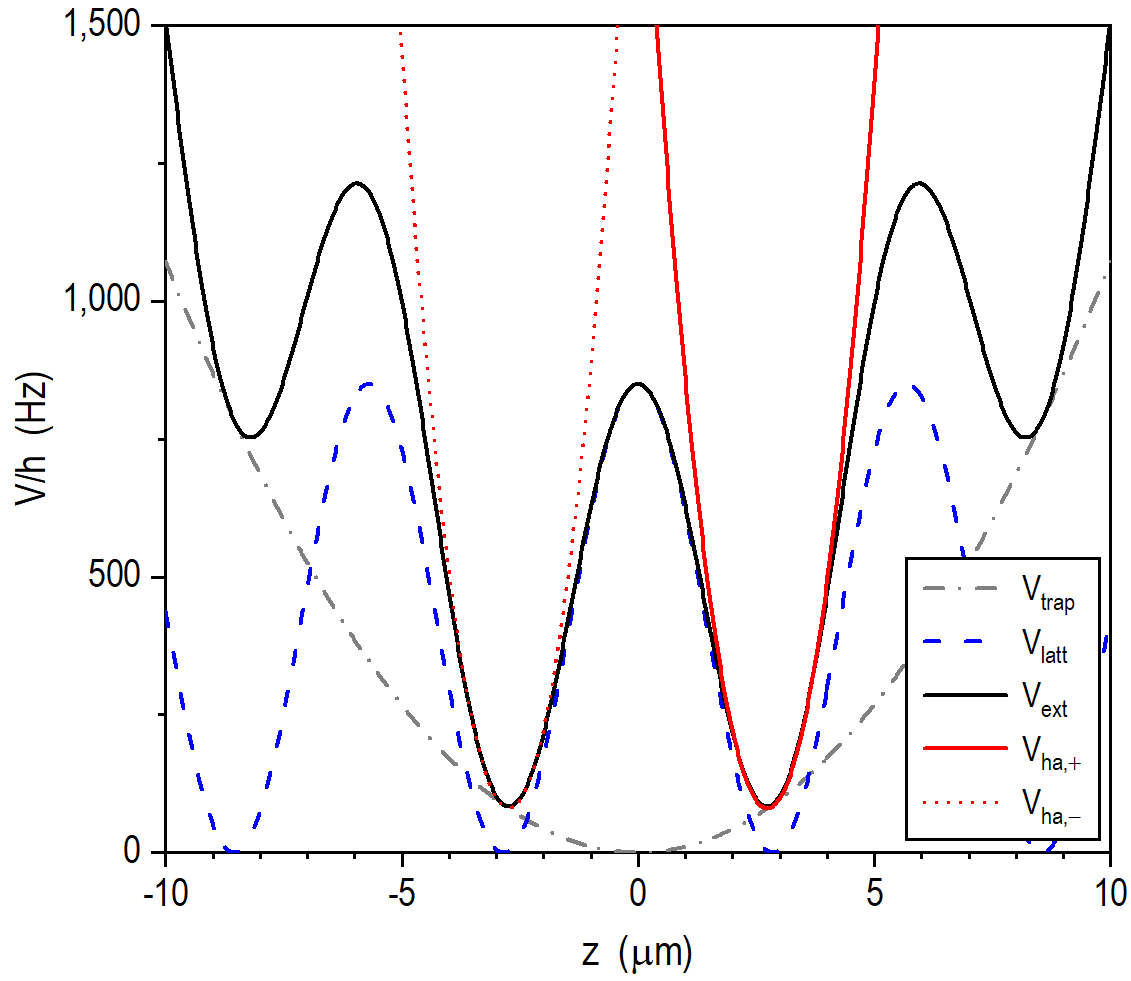}
 \caption{\label{fig1} 
  External potential (solid black line) applied to split up a single atomic condensate and generate
  a coherent superposition of two condensates.
  The confining harmonic trap (dash-dotted gray line) here has a frequency $f_z = 50$~Hz, while
  the lattice standing field (dashed blue line) has a period $\ell = 5.7$~$\mu$m and a potential
  barrier $V_0/h = 850$~Hz.
  Each site or well on either side of the central barrier can be approximated by a harmonic
  potential (dotted and solid red lines) with an effective frequency $f_{\rm eff} \approx 245$~Hz
  (see text for details), which can be used to determine the initial ansatz in the simulations.
  }
\end{figure}

The confinement of neutral cold atoms in periodic lattices of reduced dimensions, which enables the investigation of BECs and their applications \cite{lewenstein:2007:AdvPhys}, including dark soliton dynamics \cite{louis:JOptB:2004}, relies on the generation of periodic optical traps by counterpropagating laser beams \cite{grimm:AdvAtMolOptPhys:2000,bloch:RMP:2008}.
Thus, let us consider that a single atomic cloud, trapped in a harmonic potential, is adiabatically split up by applying an optical lattice a standing laser field.
If the steepness of the harmonic trap is relevant in relation to the period of the optical lattice and the depth of its wells (see Fig.~\ref{fig1}), and we assume an unbiased (or nearly unbiased) splitting of the cloud, we can assume that the BEC breaks into two, one half of each occupying a local minimum of the trap+lattice potential.
To better understand the process in terms of the potential generated (for further details on these optical lattice potentials, see, for instance, Ref.~\cite{grimm:AdvAtMolOptPhys:2000,bloch:RMP:2008}), and hence the model here considered, let us consider that the confining potential along the $z$-direction consists of two contributions, namely,
\be
 V_{\rm ext}(z) = V_{\rm trap}(z) + V_{\rm latt}(z) ,
 \label{pot1}
\ee
where the harmonic trapping contribution reads as
\be
 V_{\rm trap}(z) = \frac{1}{2} m \omega_z^2 z^2 ,
 \label{pot2}
\ee
and the optical lattice as
\be
 V_{\rm latt}(z) = V_0 \cos^2 \left( \frac{\pi z}{\ell} \right) .
 \label{pot3}
\ee

Here we consider a $^{85}$Rb atomic cloud, with $m = 1.44 \times 10^{-25}$~kg,
so we have chosen typical experimental values
\cite{ronzheimer:PhDThesis:2008,weller:PhDThesis:2009}
for the parameters involved in these two contributions, in particular,
$f_z = 50$~Hz ($\omega_z = 2\pi f_z \approx 314.2$~rad/s) and $V_0/h = 850$~Hz (where
$h = 6.62607 \times 10^{-34}$~J$\cdot$s is Planck's constant).
Furthermore, with the purpose to discuss different dynamical behaviors, three values of the
lattice constant, $\ell$, will be considered below: 5.7~$\mu$m, 15~$\mu$m, and 26~$\mu$m.
To illustrate the overall effect, the two separate contributions as well as the combined
potential (\ref{pot1}) are displayed in Fig.~\ref{fig1} for $\ell = 5.7$~$\mu$m.
The trap potential is denoted with the gray dashed line and the lattice potential with the
blue dashed line, which their sum is represented with the black solid line.
As it can be noticed, the activation of the lattice potential generates a barrier that is
going to split up the atomic cloud basically into two parts.
In a first approximation, each one of these parts can be assumed to be acted by a harmonic
well with its frequency being determined by the lattice properties (barrier height and
period).
As it can readily be noticed, in a good approximation, for the case displayed in
Fig.~\ref{fig1}, these harmonic well can be approximated by the functional form
\be
 V_{\rm ha, \pm}(z) \approx V_{\rm trap}(z_\pm)
   + \frac{1}{2} m \omega_{\rm eff}^2 (z - z_\pm)^2 ,
 \label{pot5}
\ee
with $z_\pm \approx \pm \ell/2$ and
\be
 \omega_{\rm eff} = \sqrt{\frac{2\pi^2 V_0}{m \ell^2}} .
 \label{pot6}
\ee
From this expression, we can choose for the initial width of the Gaussian wave packet the one
corresponding to the ground state of the harmonic oscillator, i.e.,
\be
 \sigma_{\rm eff} = \sqrt{\frac{\hbar}{2m\omega_{\rm eff}}} .
 \label{pot7}
\ee
Of course, a more precise (though still approximated) form can easily be derived.
However, for the purpose here, expression (\ref{pot6}) suffices, since we are interested in
releasing two coherently separated clouds from a fixed distance, $\ell$, which are assumed
to be kept inside harmonic wells determined by the lattice potential (\ref{pot3}) and not
from the combined effect (\ref{pot1}).
Note that, as $\ell$ increases, the corresponding neighboring wells become shallow very
rapidly, thus loosing their appearance of a harmonic well, while the separation between
the corresponding minima increases relatively slowly (below the value of $\ell$).


\subsection{Numerical details}
\label{sec23}

In the calculations below, we have considered a cloud of 950 $^{87}$Rb atoms, with
scattering length $a_s = 90\, a_0 = 4.76$~nm \cite{pethick2008bose}.
With respect to the coupling constant $g_{\rm 1D}$ that appears in the 1D GPE, we have considered a transverse trapping frequency value within the range indicated above, in particular, $f_\perp = 408$~Hz.
With this value, $a_\perp = 534.5$~nm and $g_{\rm 1D} = 2.45 \times 10^{-36}$~Kg$\cdot$m$^3$/s$^{-2}$.
Given the disparity of orders of magnitude involved in the different variables and parameters, in the calculations, for computational convenience, we have rescaled them in the 1D GPE (\ref{GPE1D}).
Specifically, first we have considered lengths in microns and times in milliseconds both in variables and in parameters (i.e., $\bar{x} = x/10^{-6}$ and $\bar{t} = t/10^{-3}$), which are typical orders of magnitude for distances and times involved in experiments with condensates \cite{ronzheimer:PhDThesis:2008,weller:PhDThesis:2009}, and then we have divided on both terms by $\hbar$, where the wave function, with dimensions of $length^{-1/2}$, is also rescaled ($\bar{\psi} = \psi/10^3$).
This leads to the following working functional form
\be
 i \bar{\hbar}\ \frac{\partial \bar{\psi}}{\partial \bar{t}} =
  \left( - \frac{\bar{\hbar}^2}{2\bar{m}}\frac{\partial^2}{\partial \bar{z}^2}
  + \frac{1}{2}\ \bar{m} \bar{\omega}_z^2 \bar{z}^2 + \bar{g}_{\rm 1D} |\bar{\psi}|^2 \right) \psi ,
 \label{GPE1Db}
\ee
with $\bar{\hbar} = 1$, $\bar{m} = 10^9 \times m/\hbar \approx 1.37$, $\bar{\omega}_z = 10^9 \times \omega_z/\hbar \approx 0.36$, and $\bar{g}_{\rm 1D} = 10^9 \times g_{\rm 1D}/\hbar \approx 23.1895$.

Without any loss of generality, we have assumed a Gaussian ansatz may fairly well describe the
initial BEC density distribution in a qualitative manner.
Note that we are interested here in the long-time dynamics rather than on the fine-grained
structure of the short-living initial distribution, which will start getting the appropriate
shape by virtue of the GPE nonlinear term.
Thus, in one dimension, the initial wave functions are given by coherent superpositions of
two identical Gaussian wave packets, with the functional form \cite{sanz:JPA:2008}
\ba
 \psi(z) & = & \psi_+(z) + e^{i\phi} \psi_-(z) \nonumber \\
  & \sim & e^{-(z - z_+)^2/4\sigma_0^2} + e^{i\phi} e^{-(z - z_-)^2/4\sigma_0^2} ,
 \label{ansatz}
\ea
where $z_\pm$ refers to the centroid position of each wave packet ($z_\pm = \pm \ell/2$),
$\sigma_0$ is their width, which can be conveniently related to the lattice, as discussed
in above, and $\phi$ is a constant phase factor that can be imprinted on one of the
condensates in order to analyze the influence of phase shifts.
Note that, physically, this phase factor plays might the same role as the imprinting phase
shift necessary to generate a dark soliton of the type described by Eq.~(\ref{solit1}),
but also a Bohm-Aharonov-type shift between the two clouds.
In both case, the same dynamical behavior is going to be observed in the evolution of the
corresponding density distribution.

All initial states are numerically normalized to unity, so the analytical expression of
the norm prefactor is irrelevant in (\ref{ansatz}) as it will depend on the grid size
(see below).
Apart from their simplicity, these ans\"atze are also convenient from an analytical viewpoint,
since their time-evolution is given by an analytical expression both in free space and inside
a harmonic potential in the linear case $g_{\rm 1D}=0$ \cite{sanz-bk-2}, two scenarios that
have been formerly considered in the literature \cite{Lee2012}.
This can be of some help to provide fair guesses regarding the choice of initial
parameters and also to interpret the simulations.

Finally, regarding the time evolution, the split-operator method has been considered
\cite{feit-fleck:ApplOpt:1978,feit-fleck:ApplOpt:1980-1,feit-fleck:JCompPhys:1982}.
In brief, taking into account a small time step $\delta t$, the action of the evolution
operator that actualizes the wave function from $t$ to $t+\delta t$ can be split up in
three separate contributions,
\be
 \hat{U}(t) = e^{-i\hat{H}\delta t/\hbar}
 \approx e^{-i\hat{K}\delta t/2\hbar} e^{-i\hat{V}\delta t/\hbar}
  e^{-i\hat{K}\delta t/2\hbar} ,
 \label{evolop}
\ee
where $\hat{K}$ denotes the kinetic part of the Hamiltonian and $\hat{V}$ gathers the
combined action of the external potential and the self-interaction nonlinear term.
The kinetic contribution is solved in the momentum space, where $\hat{K}$ is diagonal;
to enhance the performance, the Fourier and inverse Fourier transforms are computed by
means of the fast Fourier transform method \cite{press-bk-2}, which is also
advantageously used to evaluate the equation of motion in terms of the corresponding
spectral decomposition (plane wave basis), and then to synthesize
the Bohmian trajectories by numerically integrating it once the initial condition has been assigned \cite{sanz-bk-2}.
The spatial part is solved by discretizing the wave function on an $N$-knot mesh, in
particular, for the cases considered below, it suffices $N=1,024$.
Regarding the time evolution, a time step $\delta t = 10^{-3}$~ms ensures good convergence
in both cases (energy and norm are both preserved).


\section{Results}
\label{sec3}


\subsection{Soliton formation with two freely released BECs}
\label{sec31}

Let us first examine the formation process of dark solitons that arise after a BEC is
first adiabatically split up by turning on the optical lattice, and then, after having
separated it into two halves, the lattice is switched off, releasing the two portions
again.
Because of the harmonic trap, both parties will start moving against each other, which
leads to the appearance of interference-like fringes after some time, even though the
1D GPE is a nonlinear equation and hence does not satisfy the superposition principle.
In order to evaluate how the size of the two halves (measured in terms of the width
$\sigma_0$ of the corresponding wave packets) influences the appearance of such fringes
(or dips in the full density distribution), let us considered that the condensates are
let to freely evolve, i.e., $\omega_z \approx 0$.
Physically, this condition can be satisfied if the two condensates are relatively close
one another, so that they do not feel an important downhill acceleration prior to their
interaction, and for times much smaller than a quarter of the classical period,
$T=2\pi/\omega_z$.

Taking into account the above conditions, it is expected that the two density distributions
will essentially remain centered at the same position for the whole propagation, although
their width will undergo an increase, larger as $\sigma_0$ becomes smaller.
In the linear case ($g_{\rm 1D} = 0$), if $\omega_z \approx 0$ and $t \ll T/4$, the time
evolution of the wave packets $\psi_\pm(z)$ is given by the expression \cite{sanz:JPA:2008}
\be
 \psi_\pm (z,t) \sim e^{-(z - z_\pm)^2/4\sigma_0 \tilde{\sigma}_t} ,
 \label{ansatzt}
\ee
where
\be
 \tilde{\sigma}_t = \sigma_0 \left( 1 + \frac{i t}{\tau} \right) ,
 \label{ispread}
\ee
with
\be
 \tau = \frac{2m\sigma_0^2}{\hbar}
\ee
being an effective time scale that determines how fast the wave packet spreads,
since its spreading is given by
\be
 \sigma_t = |\tilde{\sigma}_t| =
  \sigma_0 \sqrt{ 1 + \left( \frac{t}{\tau} \right)^2 } .
 \label{rspread}
\ee
For instance, for $\sigma_0$ given by Eq.~(\ref{pot7}), if $\ell = 5.7$~$\mu$m,
we obtain $\tau \approx 1.865$~ms, which means that only after a time much larger
than this characteristic time scale, we will start observing an important overlapping
between the two wave packets, and hence the appearance of interference features.
Since the superposition principle can be applied in this case, the total wave function
is simply
\be
 \psi(z,t) = \psi_+ (z,t) + e^{i\phi} \psi_-(z,t) ,
 \label{ansatztt}
\ee
and the density distribution will be
\ba
 n(z,t) & = & n_+(z,t) + n_-(z,t) \nonumber \\
  & & + 2 {\rm Re} \left[ \psi_+(z,t) \psi^*_-(z,t) e^{-i\phi} \right ] ,
 \label{denst}
\ea
with $n_\pm (z,t) = |\psi_\pm (z,t)|^2$.
Taking into account Eqs.~(\ref{ansatzt}) and (\ref{ispread}), the interference term in
(\ref{denst}), which we will denote from now on as $n_I(z,t)$, reads as
\be
 n_I(z,t) \sim 2 e^{-[z^2 + (\ell/2)^2]/2\sigma_t^2}
  \cos \left( \frac{\hbar t \ell}{4 m\sigma_0^2} \frac{z}{\sigma_t^2}
   - \phi \right) .
 \label{densI}
\ee
When the spread of the two Gaussian distributions is such that they overlap importantly,
a rather suitable approximation for Eq.~(\ref{denst}) is
\be
 n(z,t) \approx 4 e^{-z^2/2\sigma_t^2}
  \cos^2 \left( \frac{\hbar t \ell}{8 m\sigma_0^2} \frac{z}{\sigma_t^2}
  - \phi/2 \right) .
 \label{denstapp}
\ee
Accordingly, leaving aside the Gaussian envelope (determined by the Gaussian prefactor), the distance between any two consecutive interference maxima or
minima is
\be
 \Delta z = \frac{8\pi m \sigma_0^2 \sigma_t^2}{\hbar t \ell} ,
 \label{dist}
\ee
which, for large $t$ compared to $2m\sigma_0^2/\hbar$ (but still much smaller than $T/4$),
simplifies to
\be
 \Delta z \approx \frac{2\pi\hbar t}{m\ell} .
 \label{dist2}
\ee
Therefore, once the Gaussian distributions have undergone an important spread, the distance
between consecutive fringes will not depend on the initial width $\sigma_0$ of the wave
packets, but only on the distance $\ell$ between their centroids.
Furthermore, notice that all interference maxima have the same width $\Delta z$.
As for the phase difference between the two wave packets,
$\phi$, it only causes a displacement of the fringes, producing a prominent
dip at $z=0$ in the particular case $\phi = \pi$.

\begin{figure}[!t]
 \centering
 \includegraphics[width=\columnwidth]{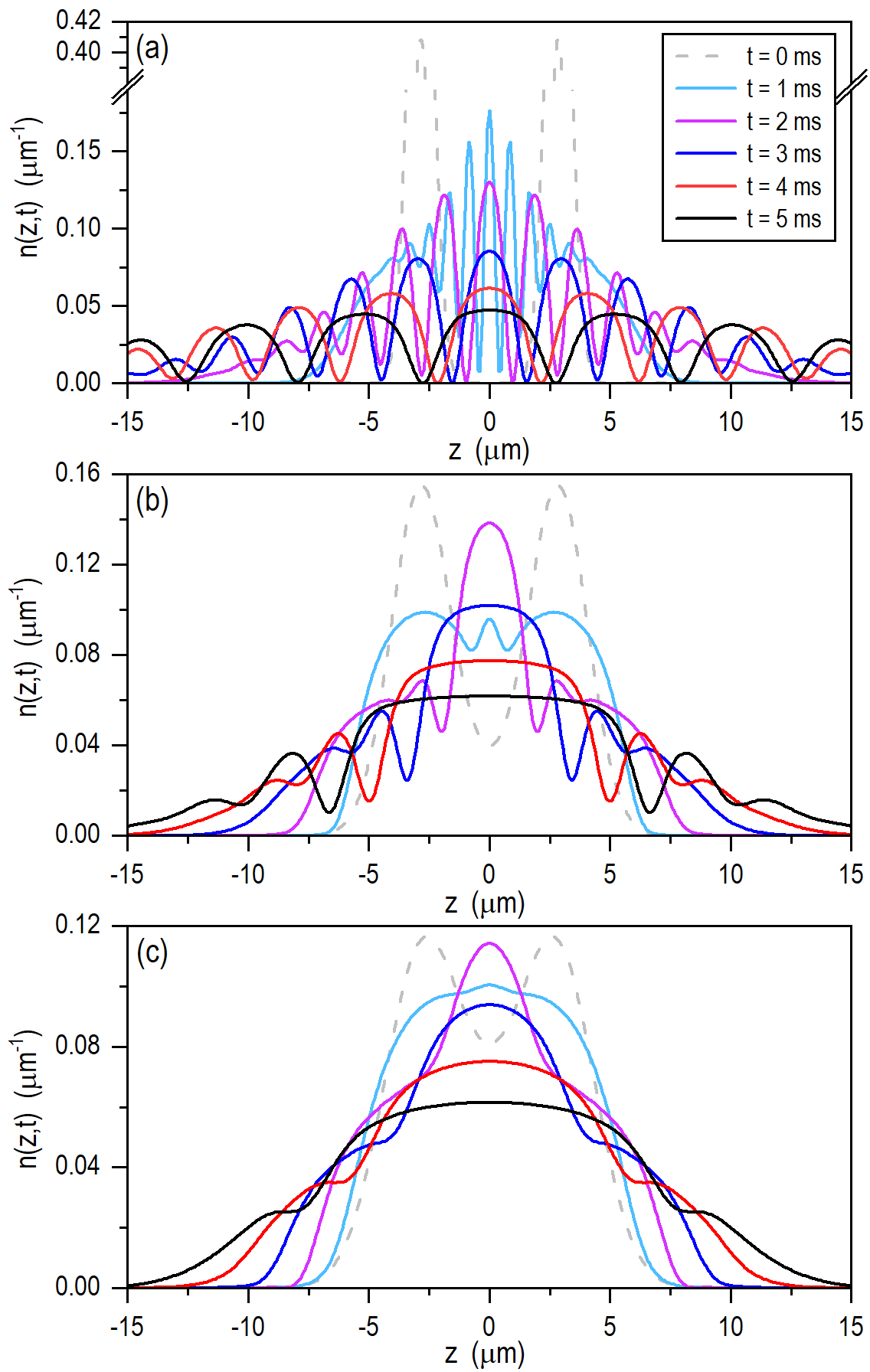}
 \caption{\label{fig2}
  Density distribution at various times (see color legend in upper panel) illustrating different
  stages of the dynamical evolution of a coherent superposition of two identical Gaussian condensates
  with an initial peak-to-peak separation $\ell = 5.7$~$\mu$m.
  In all panels, the width of the initial Gaussian is specified by the expression
  $\sigma_0 = r \sigma_{\rm eff}$, where $\sigma_{\rm eff} \approx 0.49$~$\mu$m is specified by Eq.~(\ref{pot7}) and $r$ is a parameter related to the initial (at $t=0$) overlapping between the densities $n_+$ and $n_-$ that appear in Eq.~(\ref{denst}).
  Here, this overlapping is measured in terms of the ratio $S \equiv n(0,0)/n_{\rm max}(z_\pm,0)$, where $n(0,0)$ is Eq.~(\ref{denst}) evaluated at $z=0$ and $n_{\rm max}(z_\pm,0)$ is the maximum value reached by $n(0,0)$ (at $z_+$ or $z_-$).
  Thus, from top to bottom: (a) $r = 1$ and $S \approx 0$, (b) $r = 2.5$ and $S \approx 0.26$, and
  (c) $r = 3.2$ and $S \approx 0.70$.}
\end{figure}

\begin{figure*}[!t]
 \centering
 \includegraphics[width=\textwidth]{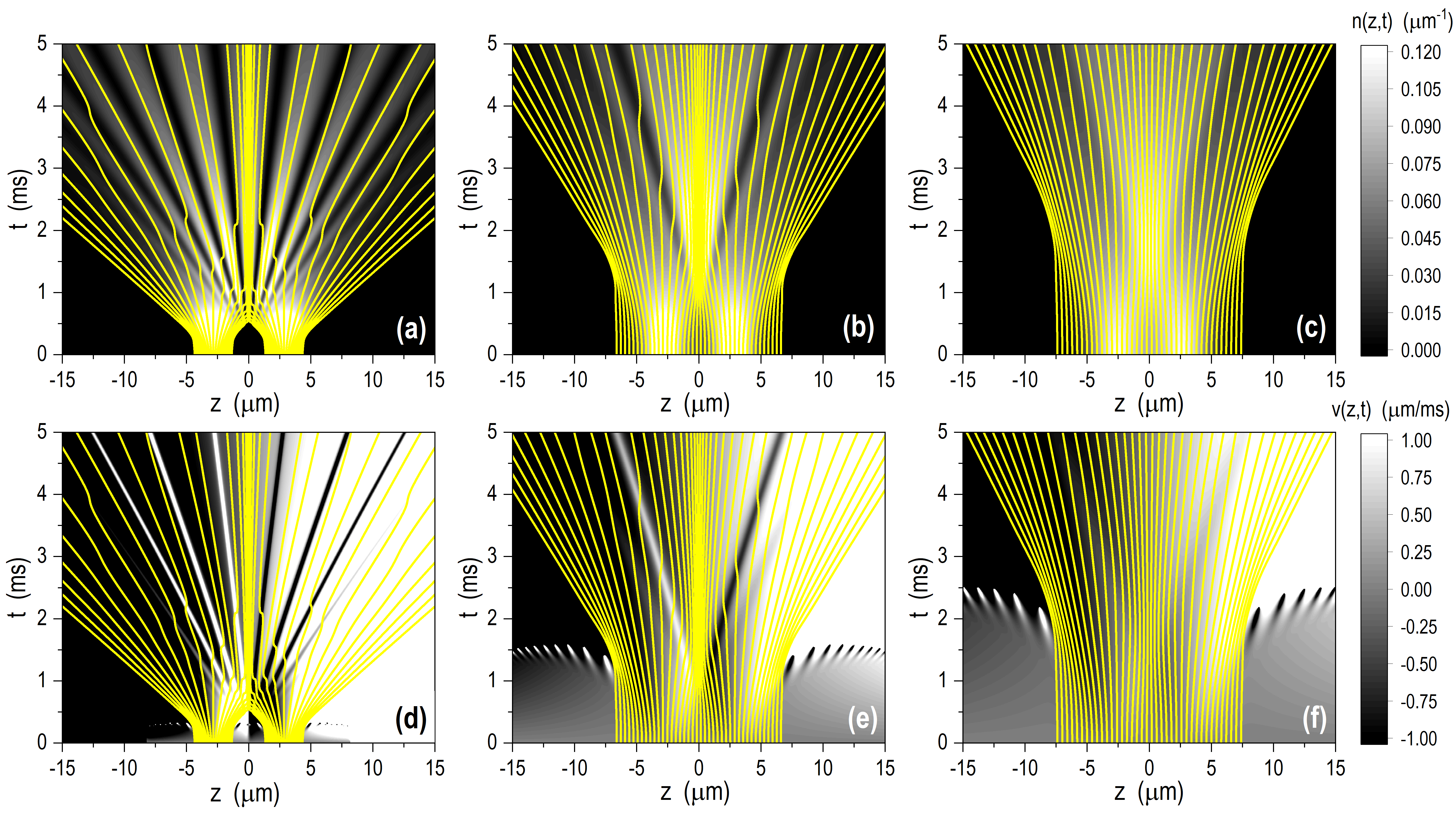}
 \caption{\label{fig3}
  Time-evolution of the BEC density distribution (top row) and its associated velocity field
  (lower row) for the three cases displayed in Fig.~\ref{fig2}: (a/d) $r = 1$, (b/e) $r = 2.5$,
  and (c/f) $r = 3.2$.
  The shading code to the right denotes the scale considered, from black for the lowest values of the
  quantities displayed to white for the highest values considered.
  For a better visualization and comparison, the density distributions have been truncated to
  0.12~$\mu$m$^{-1}$, while the velocity field ranges within $\pm 1$~$\mu$m/ms.
  To illustrate the flux dynamics, in all plots sets of 50 Bohmian trajectories (solid yellow lines)
  have been superimposed, with their initial conditions covering homogeneously regions of the density
  distribution larger than 0.5$\%$ its highest value.
  (For a better visualization of only the density plots, they are provided without the sets
   of trajectories in the form of Supplemental Material; see Fig.~S3.)}
\end{figure*}

Based on the above analytical results, let us now investigate what happens in the nonlinear
case, when $g_{\rm 1D} \ne 0$.
In Fig.~\ref{fig2}, in each panel, we show the density profile at three different times for
an initially coherent superposition of two Gaussian BECs with different widths, but all of
them with their centroids separated the same distance $\ell = 5.7$~$\mu$m.
The value of $\sigma_0$ increases from top to bottom in order to show the effect of the
nonlinearity on the overlapping clouds (i.e., on densities that have not been efficiently
separated).
As it can be readily noticed, although $\ell$ is the same in all cases, the nonlinear term
gives rise to a different behavior regarding the appearance of interference-like traits.
Thus, as $\sigma_0$ increases, from top to bottom in the figure, we observe that the minima
in the merging single condensate become less shallow, acquiring the shape of a slight
perturbation on the general profile [see panel (b)].
It is also remarkable that, unlike the linear case, where the three cases should exhibit
maxima and minima evenly spaced (at least in those cases with a high visibility, which
here correspond to small $\sigma_0$), the central maximum is wider than the side ones,
although of the order of the value rendered by Eq.~(\ref{dist2}),
which here is about 5~$\mu$m for $t=5$~ms [in agreement with what we can observe in Fig.~\ref{fig2}(a), black curve].

\begin{figure*}[!t]
\centering
 \includegraphics[width=\textwidth]{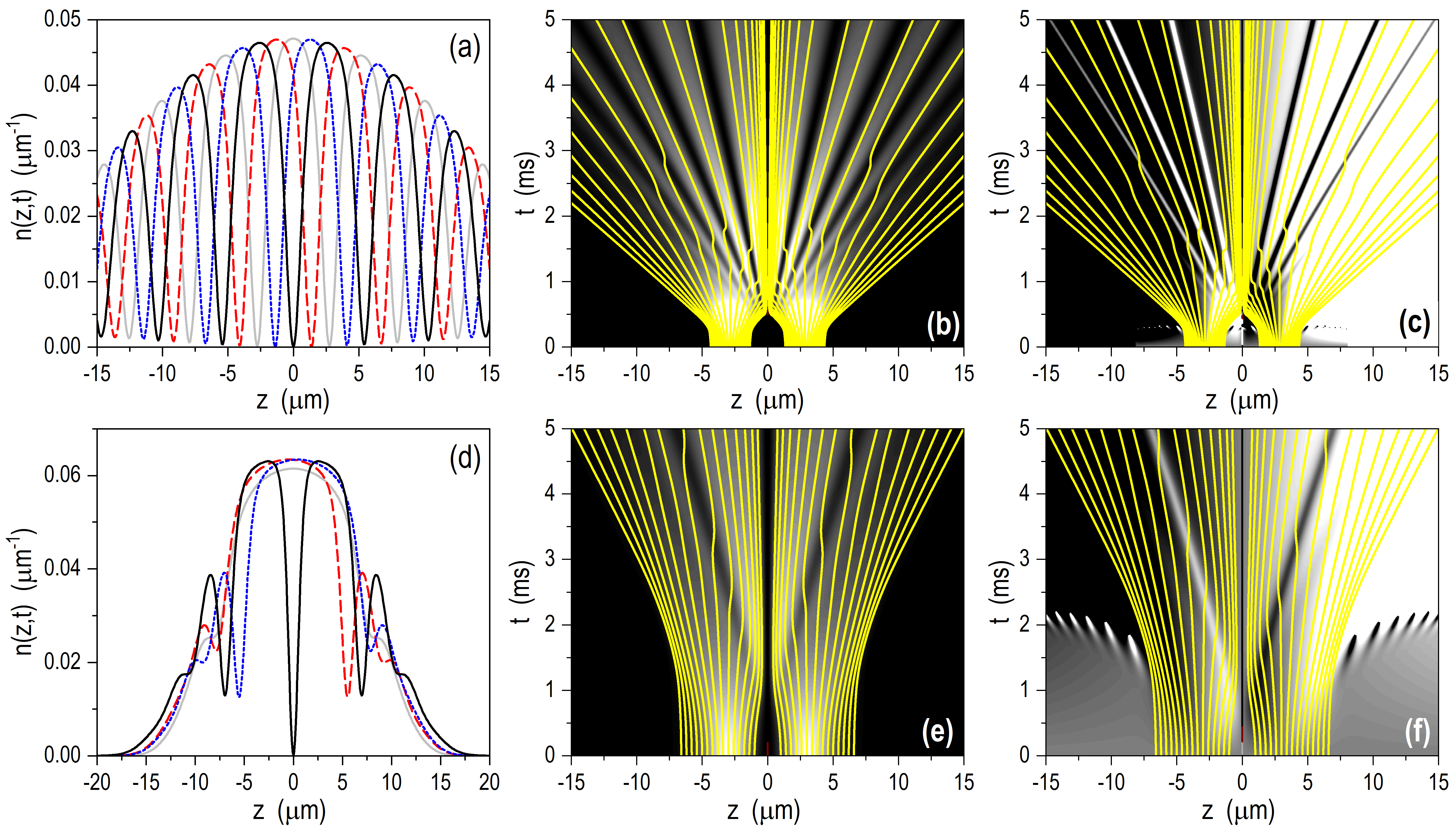}
 \caption{\label{fig4}
  Effect of the phase difference $\phi$ on the dynamics exhibited by a coherent
  superposition of two condensates with identical Gaussian distributions, with peak-to-peak
  distance $\ell = 5.7$~$\mu$m and width $\sigma_0 = r \sigma_{\rm eff} \approx 0.49 r$~$\mu$m.
  In the upper row, results for $r = 1$: (a) density profiles at $t = 5$~ms for $\phi = 0$
  (solid gray line), $\pi/2$ (dashed red line), $\pi$ (solid black line), and $-\pi/2$
  (dotted blue line); (b) and (c) density plots showing the time evolution of the density
  distribution and the velocity field, respectively, together with the corresponding Bohmian
  trajectories superimposed (solid yellow lines) for $\phi = \pi$.
  In the lower row, the same for $r=3.2$.
  The shading code for panels (b), (c), (d), and (f) is the same as in Fig.~\ref{fig3}.
  (For a better visualization of only the density plots, they are provided without the sets
   of trajectories in the form of Supplemental Material; see Fig.~S4.)}
\end{figure*}

The dynamics giving rise to the appearance of the interference-like traits discussed
above can be better understood by inspecting the density plots shown in Fig.~\ref{fig3},
where the evolution of the three superpositions is monitored until $t=3$ by means of
Bohmian trajectories (see yellow solid lines superimposed in each panel).
For a better understanding of the BEC hydrodynamics, equidistant initial conditions have
been considered, spanning a region well beyond very low values of the density.
Thus, comparing these trajectories with those associated with a linear case
(see, for instance, \cite{sanz:FrontPhys:2019}), here there is a more abrupt turn outwards
of the trajectories (flow) once the two clouds start overlapping, although at a qualitative
level in both the linear case and the nonlinear one similar trends are observable.
By inspecting the upper row in Fig.~\ref{fig3}, from left to right, we observe how the
interference-like structure gets diluted as the initial overlapping of the two condensates
becomes larger (with increasing $\sigma_0$).
On the other hand, the lower row, which shows the density plot of the evolution in time of
the (drift) velocity field provides us with an alternative picture for such a behavior.
Although initially this field is zero, because there is no local variation of the phase
in the initial wave function, for low values of $\sigma_0$, after a very short time, the
velocity field acquires important
negative (dark shading, from gray to black; see shading code in the
r.h.s.\ legend) and positive (faint shading, from gray to white) values,
dividing the space dynamically.
A first division is related to the symmetry line, at $x=0$, which will remain for all times;
accordingly, trajectories related to one of the clouds cannot penetrate the region determined
by the other cloud, and vice versa.
This happens for all values of $\sigma_0$ and is related to the so-called non-crossing
property of Bohmian trajectories \cite{sanz:JPA:2008}, which arises from the single-valuedness
of the underlying velocity field \cite{sanz:FrontPhys:2019}.
A second division is connected to the (space-dependent) phase acquired by each individual
cloud, which in the linear case is given by (\ref{ansatzt}), and that makes the associated
trajectories to move in opposite directions, thus causing their dispersion and hence the
widening of the density distribution.
Now, in the case of low $\sigma_0$ values, there is a longer range of interaction between
the two clouds, which results, at later times, on solitons with a higher visibility.
Note in Fig.~\ref{fig3}(d) that, between $t \approx 0.1$~ms and $t \approx 1.2$~ms, the region
between the centers of the two condensates is dominated by two intense velocity fields
with opposite values.
When $\sigma_0$ increases and the overlapping is more important, such a region decreases
and gets smoother, thus leading to fainter interference minima, even though at later times,
as seen in Fig.~\ref{fig2}(b), they become more apparent.
However, beyond certain values of $\sigma_0$, the active region between the two clouds
becomes irrelevant and interference traits blur, as seen in Fig.~\ref{fig3}(f) [see also
Fig.~\ref{fig2}(c)].

Unlike the linear case, though, a series of kinks appear in the trajectories as they become
abruptly deflected outwards (in the linear case, such deflection takes place more gradually).
Typically, these kinks appear around nodes of the wave function in the linear case, where the
velocity field undergoes fast changes in the form of positive or negative spikes (see, for
instance, Ref.~\cite{sanz:FrontPhys:2019}).
A close inspection at the lower row panels in Fig.~\ref{fig3} shows that also in the nonlinear
case this is the case, since a series of sudden and highly localized changes in the velocity
arise in the form of chains on both sides and for all values of $\sigma_0$.
These features, which are very prominent in the plots of the velocity fields, because any sudden phase change affects importantly the local velocity, can also be found in the density plots of the probability density (not shown here), but only if one considers its logarithmic representation, since they take place in regions where the density values are already very low.
On the contrary, if we launch trajectories with initial conditions in the regions where the initial density $n(z,0)$ is already negligible, they will be able to detect these important phase variations even if the density values are very low.
This is precisely where the strength of the trajectory or, analogously, velocity field method relies on: phase changes in regions with negligible densities can be immediately detected, because of the tight connection between phase correlation and Bohmian trajectory.

Given the high sensitivity of the trajectories to phase changes, they can also be used to
detect other phase-based effects.
Consider, for instance, an Aharonov-Bohm type shift, like the one induced on the
interference of two coherent electron beams acted by a shielded magnetic field, as
reported by Tonomura {\it et al.}~\cite{tonomura:PRL:1986}.
The action of the shielded field can be simulated by simply shifting the relative phase
between the electron beams.
In the linear case, a trajectory simulation of this fundamental result was already provided
by Philippidis {\it et al.}~\cite{philippidis:NuovoCim:1982}.
To some extent, due to the linearity of Schr\"odinger's equation, this is an expected result,
since the phase factor is going to directly appear in the argument of the cosine that rules
the interference term (\ref{densI}).
In the nonlinear case, though, considering $\phi \ne 0$ in (\ref{ansatztt}) does not warrant
a priori a subsequent effect at later times.
However, as in the counter-intuitive case of finding an interference-like behavior, also
such an additional phase difference between the two merging clouds is going to have an
important manifestation in the behavior of the interference traits, as shown in
Figs.~\ref{fig4}(a) and (d), for $\sigma_0$ with $r=1$ and $r=3.2$, respectively, and where
in both cases $\phi$ takes the values $-\pi/2$ (blue solid line), $\pi/2$ (green solid line),
and $\pi$ (red solid line).
To compare with, the case $\phi = 0$ (gray dashed line) has also been added in each case.
As it is seen in Fig.~\ref{fig4}(a), $\phi$ only produces a shift of the solitons, leftwards if
it is negative and rightwards if it is positive; in the case $\phi = \pi$, what we observe is
that solitons appear in the positions corresponding to the maxima of the density distribution,
and vice versa, which is due to the fact that, at $t=0$, the density has to be cancel out at
$z=0$, according to the relation
\ba
 n(0,0) & = & 2 n_\pm(0,0) + 2 n_\pm (0,0) \cos \phi \nonumber \\
  & = & 4 n_\pm (0,0) \cos^2 (\phi/2)
 \label{denst0}
\ea
(note that, at $z=0$, both $n_-$ and $n_+$ acquire the same value, so it does not matter which one
is chosen).
Note that, even if this relation is not satisfied anymore along the evolution of the condensate,
because of the lack of linearity of the 1D-GPE, the conservation of the flux constrains (made
evident through the Bohmian trajectories) implies that $n(0,t)$ must be zero at any subsequent
time.
This is precisely what we observe in Fig.~\ref{fig4}(a), where $t=5$~ms and despite of initially
having two well separated clouds.
This effect, though, is more interesting in the case of overlapping condensates.
As seen in Fig.~\ref{fig4}(d), although for $\phi = 0$ no solitons emerge, as $\phi$ is increased
(or decreased towards negative values), the start emerging, because the overlapping between the two
clouds is going to increase.
The most remarkable case takes place for $\phi = \pi$: although the two clouds are wide enough to
spatially overlap, the appearance of a $\pi$-phase difference between both will make the initial
density to vanish at $z=0$, thus generating two initially separate distributions.
As time proceeds, the same interference-like phenomenon will then be observed, which eventually leads
to the appearance of solitons, just like in the case for $\phi = 0$, although in a lesser number.
The density plots in panels (b) and (e) for the density distribution, or in panels (c) and (f) for
the local velocity field, provide us with a clear picture on how this interference process mediated
by a initial phase difference takes place, and how the flux, described by means of Bohmian trajectories,
shows the appearance of the corresponding dark solitons by a recast of the density in some regions,
while quickly avoids others (those occupied by the solitons).
These numerical simulations are in good agreement with those in
Ref.~\cite{ronzheimer:PhDThesis:2008}, where a potential energy difference of the initial
condensates is used to induce the phase difference.


\subsection{Soliton formation from two BECs released in a harmonic trap}
\label{sec32}

Let us now consider the case where the effects of the harmonic trap are important and
the two clouds are let to evolve inside, each released from an opposite site of the
potential.
As before, the distance between their centroids is $\ell$.
Regarding the initial width, also as before, we make some considerations based on the
linear case.
Thus, when a Gaussian wave packet is released inside a harmonic well with frequency
$\omega_z$, if its initial width is $\sigma_0^2 = \hbar/2 m \omega_z$, the time evolution
of the wave function is given by \cite{sanz:cpl:2007}
\ba
 \psi(z,t) & \sim &
  e^{-(z - z_0 \cos \omega_z t)^2/4\sigma_0^2 - i\omega_z t/2} \nonumber \\
  &  & \quad \times
     e^{- im\omega_z (4z z_0 \sin \omega_z t - z_0^2 \sin 2\omega_z t)/4\hbar} ,
 \label{ansatzt2}
\ea
which shows that the width of the probability density remains constant in time.
This means that the corresponding Bohmian trajectories [obtained from the phase of
Eq.~(\ref{ansatzt2})] will exhibit a recurrent oscillatory motion, according to the
equation
\be
 z(t) = z(0) - z_0 + z_0 \cos \omega_z t ,
 \label{traosc}
\ee
which also shows that the trajectories are parallel one another.
Otherwise, if the initial width is not constrained to the above value, the wave packet
undergoes a sort of ''breathing'', with its width oscillating as it moves from one turning
point to the opposite one, and vice versa.
Consequently, the corresponding trajectories will not be parallel, but will separate from
the central one (the one with its initial condition coinciding with the position of the
initial centroid of the wave packet) and can back again periodically.
If a coherent superposition of two of such wave packets is considered, each one located on
opposite turning points initially, and regardless of the value of $\sigma_0$, the associated
trajectories will undergo a bounce backwards whenever they reach the center of the potential,
as they must satisfied the above mentioned non-crossing rule \cite{sanz:JPA:2008,sanz-bk-2}.

In the present case, given that the two initial wave packets arise from a former
confinement in the wells formed by the lattice sites, we are going to assume that both
correspond to the ground state of the corresponding approximated harmonic potentials.
Thus, the superposition will take the form of (\ref{ansatz}), with $z_\pm = \pm \ell/2$
and $\sigma_0^2 = \hbar/2m\omega_{\rm eff}$.
As for the effective frequency $\omega_{\rm eff}$, we will consider Eq.~(\ref{pot6}),
which decreases with $\ell^{-1}$, since increasing $\ell$ implies wider and wider
effective harmonic wells.
Here we are going to consider the same three values for $\ell$, which make
$f_{{\rm eff},1} = 245.3$~Hz for $\ell = 5.7$~$\mu$m, $f_{{\rm eff},2} = 93.2$~Hz for
$\ell = 15$~$\mu$m, and $f_{{\rm eff},3} = 53.8$~Hz for $\ell = 26$~$\mu$m, with
$f_{\rm eff} = \omega_{\rm eff}/2\pi$.
If the nonlinear case shows some similarities with the linear one, we should expect the
appearance of different dynamical behaviors depending on whether $f_{\rm eff}$ approaches
the value of $f_z = 50$~Hz or diverts from it.
Actually, taking into account that the spreading of the wave packet under linear
conditions, given by Eq.~(\ref{rspread}), is ruled by the effective width at $t = T/4$,
$\sigma_{\pi/2} = \hbar \omega_{\rm eff}/2m\omega^2$, there will be an additional
constraint involved in the process.
Thus, we have $\sigma_1 = 2.39$~$\mu$m for $\ell = 5.7$~$\mu$m, $\sigma_2 = 1.47$~$\mu$m for
$\ell = 15$~$\mu$m, and $\sigma_3 = 1.19$~$\mu$m for $\ell = 26$~$\mu$m, to be compared
with the width corresponding to a coherent wave packet,
$\sigma_c = \hbar/2m\omega^2 = 1.08$~$\mu$m (with $f_z = 50$~Hz).

\begin{figure*}[!t]
\centering
 \includegraphics[width=\textwidth]{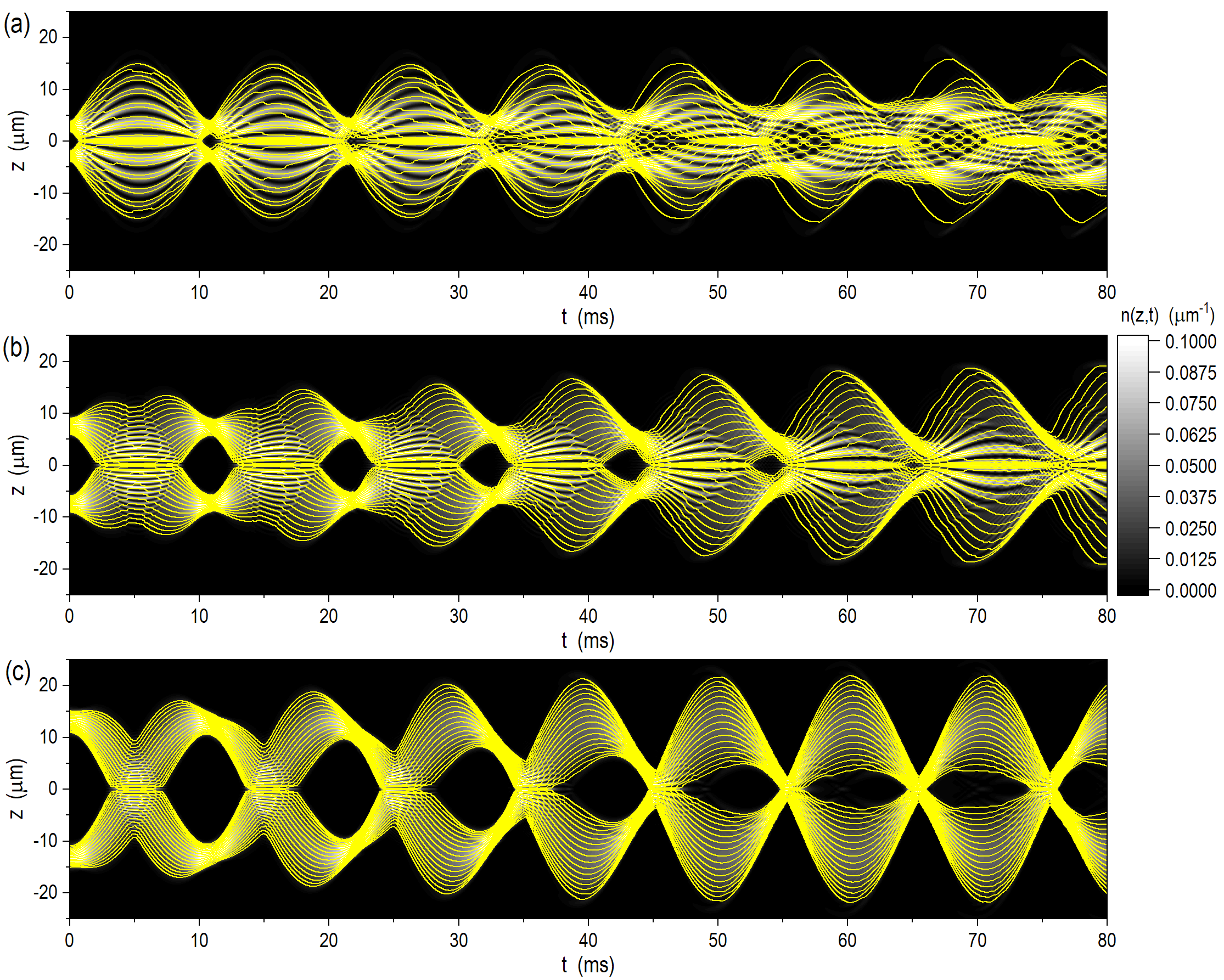}
 \caption{\label{fig5}
  Contour plots showing the time evolution of the density distribution associated
  with three initial superpositions with peak-to-peak distances:
  (a) $\ell = 5.7$~$\mu$m, (b) $\ell = 15$~$\mu$m, and (c) $\ell = 26$~$\mu$m.
  The shading code is defined on the right; the distributions have been truncated
  to $n_{\rm max} = 0.1$~$\mu$m$^{-1}$ in all cases for a better visualization.
  Sets of Bohmian trajectories (solid yellow lines) with homogeneously distributed
  initial conditions are superimposed in order to illustrate the local aspects
  exhibited by the condensate dynamics.
  In all cases the frequency of the harmonic trap is $f_z = 50$~Hz.
  (For a better visuallization of only the density plots, they are provided without the sets
   of trajectories in the form of Supplemental Material; see Fig.~S5.)}
\end{figure*}

In Fig.~\ref{fig5} we can observe the three possible dynamical scenarios that can take
place inside the harmonic trap depending on the value of $\ell$, which rules both the
separation between the two clouds (distance between two neighboring sites) and their initial
width (related to the approximated harmonic well).
Following a linear dynamics, we observe that, for $\ell = 5.7$~$\mu$m, the initial width is
quite different from the width expected for a coherent state in the harmonic trap.
Hence, it is expected that it will undergo the above mention ``breathing''.
In Fig.~\ref{fig5}(a) we observe that, effectively, because the width of the initial clouds
do not match that of a coherent wave packet, they undergo an important spreading, giving rise
to the appearance of interference-like traits at about $t = T/4$.
As it is shown by means of the superimposed trajectories, this structure repeats a few times
until it starts getting blurred precisely at the times where one would expect a revival of
the initial state, at about multiples of half the period.
A closer inspection shows that this blurring consists of a mesh of nodes that, as time
proceeds, spread all over the place that should be occupied by interference traits, which
are pushed beyond them, as it can be seen in the subsequent expansions after 30~ms.
The messy behavior of the trajectories indicates that the density between two consecutive
solitons is being transfer from the most external parts of the condensate to the innermost
ones.
It is this effective transfer what eventually leads, when it gets in contact with the flux
associated with the core part of the density, to the appearance of the web of nodes.
A different behavior is observed for $\ell = 26$~$\mu$m, as it is shown in Fig.~\ref{fig5}(c).
In this case, the effective frequency is pretty close to the frequency of the harmonic
trap, so the initial width of the clouds is close to that of coherent wave packets.
Hence, the first stages in the evolution of the clouds are analogous to those that
can be observed in the linear case, with the typical bounce backwards of the trajectories
at $t = T/2$.
It is precisely in this region where a series of aligned solitons arise, remaining for
longer times, although the nonlinearity causes that the clouds sent towards the turning
points will increase their width.
These two behaviors are pretty similar to the Young-type and collision-type interference
process that can be found when we are dealing with the linear evolution of a coherent
superposition of two wave packets \cite{sanz:JPA:2008}.
In between, as seen in Fig.~\ref{fig5}(b), we observe a transition dynamics that keeps
features from the two regimes.

\begin{figure*}[!t]
 \centering
 \includegraphics[width=\textwidth]{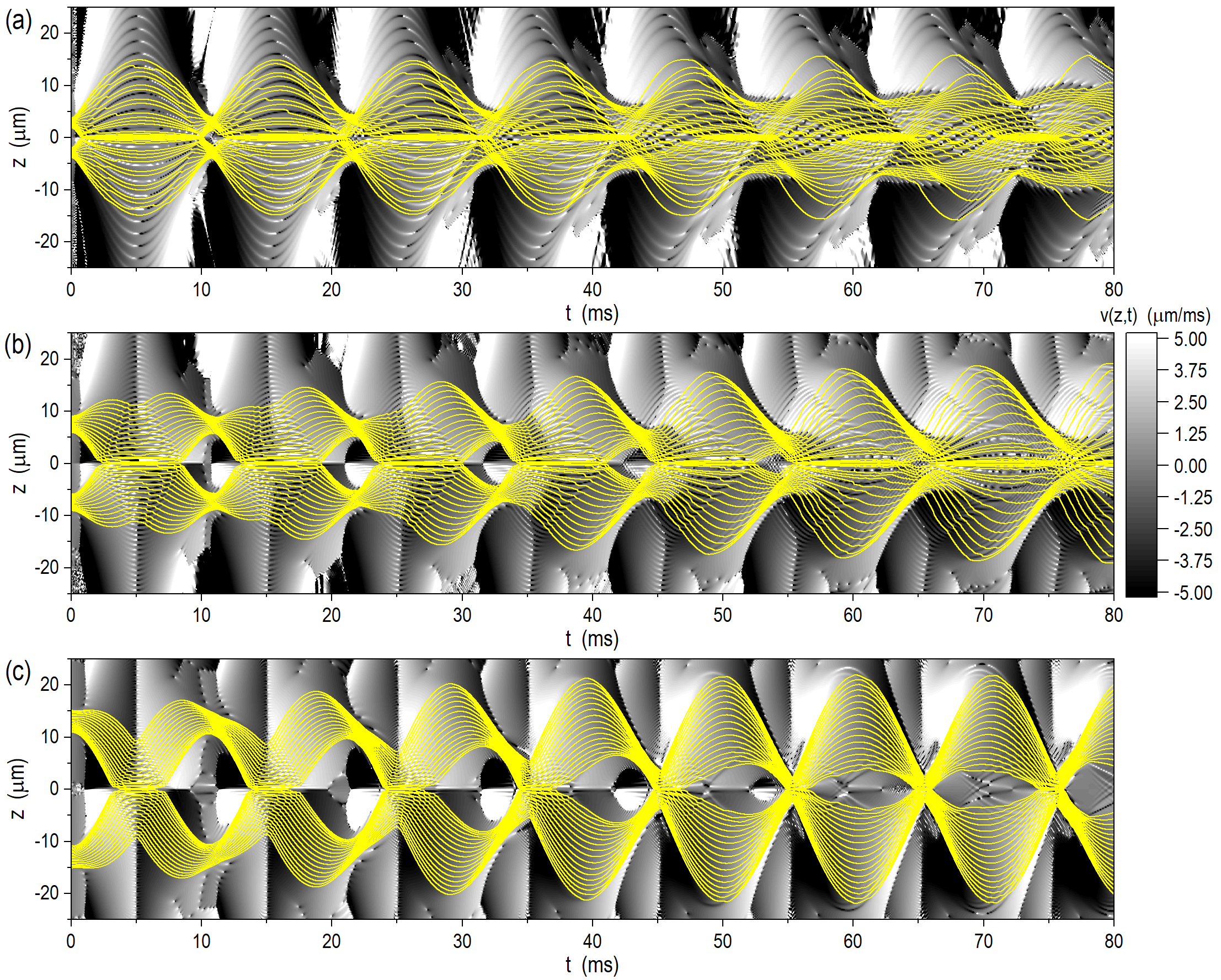}
 \caption{\label{fig6}
  Contour plots showing the time evolution of the velocity field associated with
  the three initial superpositions considered in Fig.~\ref{fig5}, with
  peak-to-peak distances: (a) $\ell = 5.7$~$\mu$m, (b) $\ell = 15$~$\mu$m,
  and (c) $\ell = 26$~$\mu$m.
  The shading code is defined on the right; velocity values have been constrained
  to the range $\pm 5$~$\mu$m/ms in all cases for a better visualization.
  Sets of Bohmian trajectories (solid yellow lines) with homogeneously distributed
  initial conditions are superimposed in order to illustrate the local aspects
  exhibited by the condensate dynamics.
  In all cases the frequency of the harmonic trap is $f_z = 50$~Hz.
  (For a better visuallization of only the density plots, they are provided without the sets
   of trajectories in the form of Supplemental Material; see Fig.~S6.)}
\end{figure*}

If we look at the corresponding velocity plots instead, displayed in Fig.~\ref{fig6}, we observe
a rather enlightening picture for the dynamical behavior exhibited by the density distribution.
As seen in Fig.~\ref{fig6}(a), for early times, the density is
acted by a velocity field that is relatively homogeneous, with positive (faint shading, from
gray to white; see shading code in the r.h.s.\ legend) and negative values (darker shading,
from gray to black).
In those regions where the velocity field acquires positive values, the trajectories (flux)
are driven upwards, towards positive $z$, while negative values lead the trajectories downwards,
along the negative $z$-direction.
At about $t=5$~ms, a series of grayish bands, representing velocity values
around 0, starts emerging, separated by other sharp negative and positive velocity regions;
the trajectories avoid these regions, accumulating along the
grayish ``valleys'', where the values of the velocity are moderate or even negligible.
At this time, there is also a broad expansion of the set of trajectories, as it corresponds to
wave packets that feel a mild effect from the confining potential; the weaker the interaction
the larger the spread range, as it happens in a linear case, in a typical Young-type two
wave-packet collinear interaction \cite{sanz:JPA:2008}.
Afterwards, because of the action of the confining potential, the swarm of trajectories starts
gathering again guided by a exchange of the sign of the velocity field (the positive region
becomes negative, and vice versa), until an approximate revival of the initial density is
observed at $t \approx 10$~ms, which corresponds to nearly half the period of the harmonic
potential, $T = 20$~ms.
Note hat, because the two condensates are identical, the trajectories cannot cross the symmetry
axis $x = 0$, and hence the repetition of the pattern takes place in half a cycle under linear
propagation conditions \cite{sanz:JPA:2008}.
The same trend is observed at subsequent times, although there is a degradation of the neat
alternating structure of the velocity field, which consequently leads to a blurring of the
region where the interferential valleys appear.
Thus, because of the presence of the nonlinearity, the two well-defined dynamical regimes,
specified by regions of opposite velocity fields and regions with alternate interference
bands, start mixing, generating new interferential patterns [see Fig.~\ref{fig5}(a)].
As it can be seen, the trajectories are very sensitive to these structures, since they
always try to avoid very quickly regions with low density, associated with sudden changes
in the velocity field.

In the opposite case, when the two wave packets start quite far apart one another, as it is seen
in Fig.~\ref{fig6}(c), the regions with opposite velocity field values are more clearly defined
and their preservation remains for a longer time.
However, the intermediate interference region with low velocity values is almost negligible, as
it would corresponds to a situation analogous to a collinear collision of two wave packets in a
linear regime \cite{sanz:JPA:2008}.
In such situations, although the trajectories cannot cross the symmetry line, it can be noticed
that, on average, they behave like moving along the trajectory of a classical harmonic oscillator.
Of course, the presence of the nonlinearity introduces the corresponding deviations, but still
we can perceive this type of back and forth motion along time.
In those regions where the two wave packets overlap (at about a quarter of a period, for
instance), the trajectories bounce backwards, but there is an effective transfer of the
state of motion that they describe from the upper swarm to the lower one, and vice versa,
which provides us with a certain sense of flow continuity.
Note that this happens by virtue of the quick velocity flipping at the instant where the two
sets of trajectories reach their maximum proximity, even at any later time, when the topology
displayed by both swarms has nothing to do with the initial swarms.

\begin{figure}[!t]
 \centering
 \includegraphics[width=\columnwidth]{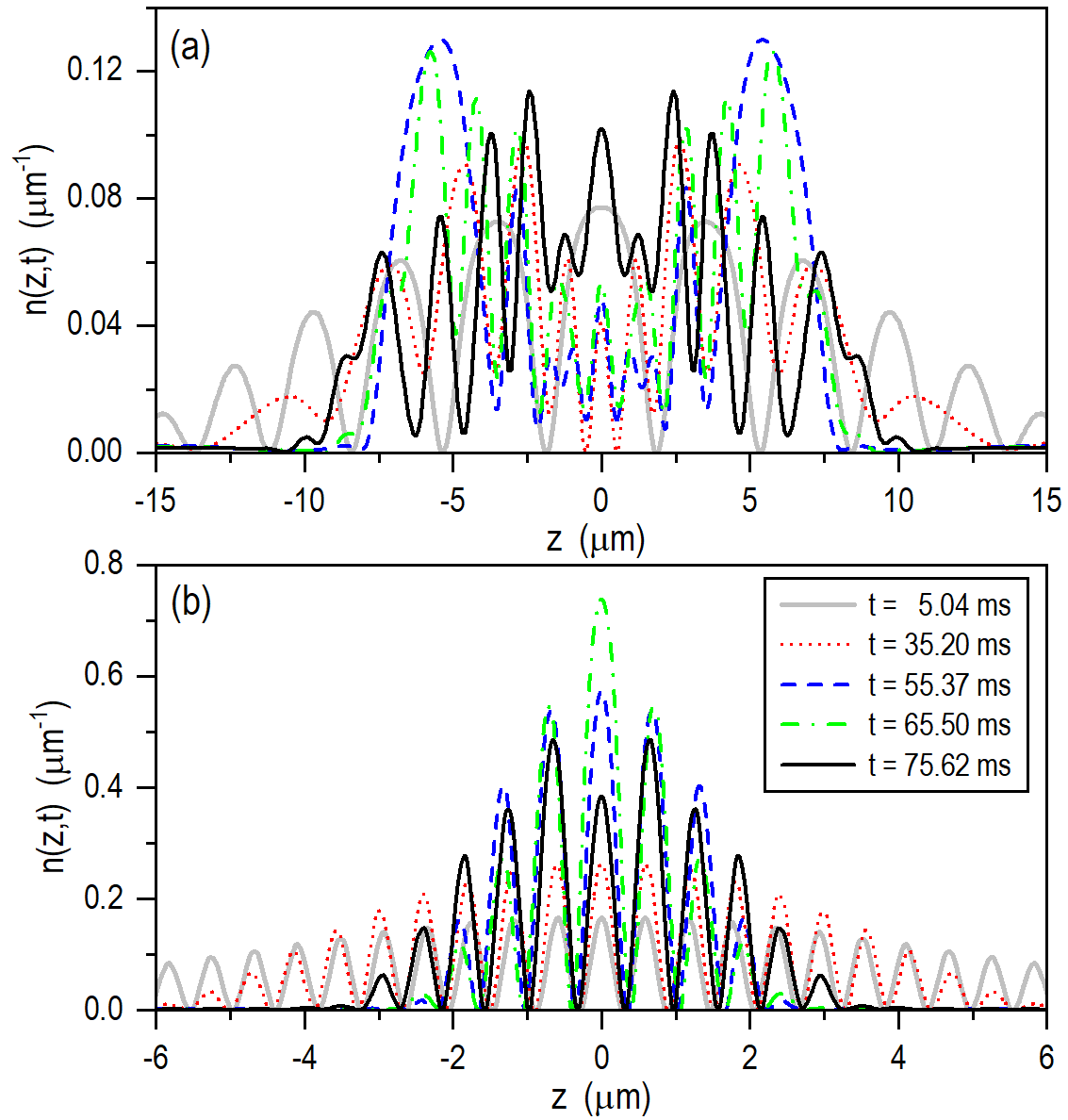}
 \caption{Density distribution at different times for coherent superpositions of two condensates
  inside a harmonic trap with $f_z = 50$~Hz and peak-to-peak distances: (a) $\ell = 5.7$~$\mu$m
  and (b) $\ell = 26$~$\mu$m.
  As it can be read in the legend inserted in the lower panel, the times considered in both cases
  are: $t = 5.00$~ms (solid gray line), $t = 35.20$~ms (dotted red line), $t = 55.37$~ms (dashed
  blue line), $t = 65.50$~ms (dash-dotted green line), and $t = 75.62$~ms (solid black line).}
  \label{fig7}
\end{figure}

The repetitive behavior of the velocity field can then be used in a profitable manner to
determine the precise times at which maximum fringe visibility is expected in the merging
of the two condensates by simply determining the time at which the velocity field flips
take place.
Equivalently, the same can be done from the times at which the sets of trajectories undergo
the bounce backward, something that cannot be unambiguously done in the case of initially
close wave packets beyond few periods.
This fact is illustrated in Fig.~\ref{fig7}, where the density distribution for $\ell = 5.7$~$\mu$m
[part (a)] and $\ell = 26$~$\mu$m [part (b)] has been plotted for some times at which there is a
sudden velocity flip in the case displayed in Fig.~\ref{fig6}(c).
For $\ell = 5.7$~$\mu$m, we observe that the series of dark solitons emerged by interference disappears
very quickly.
For $\ell = 26$~$\mu$m, however, although the ranged occupied by the series of dark solitons decreases
with time, they still keep showing up with almost the same width and period.
Accordingly, we find that the process of dark soliton generation by merging two condensates, of much
interest for BEC interferometry, is thus more robust as such condensates are more distant initially,
i.e., as the period of the lattice becomes larger.

\vspace{-.25cm}
\section{Concluding remarks}
\label{sec4}
\vspace{-.25cm}

The dynamics involved in matter-wave interferometry with condensates
are known to be well described by the mean-field theory in terms of the GPE.
Despite of the nonlinearity of the GPE, the coherent merging
of two formerly split up condensates gives rise to the appearance of
interference-type structures, where the minima can be identified with
dark solitons, which, in turn, arise as a consequence of the atom-atom
repulsion (implicitly included in the GPE nonlinear term).
In order to provide an alternative perspective on the dynamical processes
where these structures arise, here different scenarios have been analyzed
within the hydrodynamical framework of Bohmian mechanics.
Thus, the velocity field associated with local phase variations
of the wave function (order parameter) and the flux trajectories that evolve
accordingly render a different view and understanding of the appearance of
interference-type features.
To stay close to known situations, in particular, here we have considered cases
previously investigated both numerically and experimentally in the literature,
characterized by the appearance of arrays of dark solitons
\cite{ronzheimer:PhDThesis:2008,weller:PhDThesis:2009}.
As it is shown, although the process somehow resembles the emergence of interference
fringes in the standard linear case (i.e., in a typical Young-type experiment),
the presence of nonlinearities invalidates the direct application of the superposition
principle and, therefore any explanation or interpretation based on it.

Following the current (mean field) hydrodynamical description, it is noticed that
the gradual appearance of the dips in the condensate density distribution, which
are commonly identified with interference minima, can be identified with the formation
of dark solitions, because they are associated with a phase difference analogous to
that involved in the imprinting technique that provokes the generation of single dark
solitons in a homogeneous cloud.
In the current case, instead of only having the usual step-type phase variation that
is imprinted on the condensate, there is an alternating (almost periodic) succession of
such step-type structures, which are detected within our approach by computing the local
velocity field (directly related to the gradient of the phase) and the flux (Bohmian)
trajectories.
At such positions, the velocity field undergoes a sudden and prominent increase or
decrease, which it remains nearly constant between any two of these ``kicks''.
The presence of these localized changes in the velocity provides us with a clue on
where a dark soliton has formed.
Furthermore, as a consequence, in these regions the flux feels a sudden acceleration,
which provokes fast turns in the corresponding trajectories.
This thus forces the trajectories to jump from a given spatial region of nearly zero
acceleration to a neighboring one, while leaving empty the region where they undergo
the boost remains lowly populated.
This process, with flux trajectories getting accumulated in certain regions, explains
the inhomogeneity in the intensity distribution, with a structure resembling that of
an interference pattern, although the dynamical origin is different.
This behavior has been observed both in freely released condensates and also in
condensates released inside a harmonic trap.

In relation to the above comment, in the case of freely released condensates, we have
seen that by simply adding a phase difference between the two condensates that constitute
the initial superposition state automatically leads to the appearance of soliton-type
features.
Of course, this can be expected, because the process is basically the same as the
imprinting technique used to generate dark solitons in a single cloud, even though
here we start with two spatially separated clouds.
However, note that, on the other hand, this situation also resembles that of the
Aharonov-Bohm effect, where a local phase change in one of the two branches of the
wave function leads to a measurable shift of the interference features.
The representation in terms of velocity fields and trajectories allow us to detect the
effect generated by these initial phase differences between the two clouds.
In particular, for $\phi = \pi$, a deep dark soliton will emerge at the center in the
case of identical initial clouds,
This region will be avoided by the corresponding flux trajectories, which will accumulate
either on one side or the other of this singular region, but never approaching the half
point, contrary to what happens with $\phi = 0$.

Finally, in the case of condensates inside a harmonic trap, we have also seen that a
knowledge of the velocity field or the corresponding flux trajectories has an intrinsic
practical interest, as it allows us to determine in a more accurate manner time scales
ruling the condensate dynamics.
The fact that the velocity field displays sudden changes (or the trajectories exhibit
clear turns) constitute an important advantage against simply considering an inspection
of the time evolution of the density distribution, since such changes are quite
unambiguous spatially (as well as the trajectory turns).
Additionally, from the interferometric point of view, this analysis also offers valuable
information about why the superposition of the two condensates shows a remarkable dependence
on the releasing position or how long we can wait until observing a degradation of the
oscillatory dynamics.


\vspace{-.25cm}
\section*{Acknowledgments}
\vspace{-.25cm}

Financial support from the Spanish Agencia Estatal de Investigaci\'on (AEI) and the European Regional Development Fund (ERDF) (Grant No.\ PID2021-127781NB-I00) is acknowledged.



\begin{thebibliography}{59}%
\makeatletter
\providecommand \@ifxundefined [1]{%
 \@ifx{#1\undefined}
}%
\providecommand \@ifnum [1]{%
 \ifnum #1\expandafter \@firstoftwo
 \else \expandafter \@secondoftwo
 \fi
}%
\providecommand \@ifx [1]{%
 \ifx #1\expandafter \@firstoftwo
 \else \expandafter \@secondoftwo
 \fi
}%
\providecommand \natexlab [1]{#1}%
\providecommand \enquote  [1]{``#1''}%
\providecommand \bibnamefont  [1]{#1}%
\providecommand \bibfnamefont [1]{#1}%
\providecommand \citenamefont [1]{#1}%
\providecommand \href@noop [0]{\@secondoftwo}%
\providecommand \href [0]{\begingroup \@sanitize@url \@href}%
\providecommand \@href[1]{\@@startlink{#1}\@@href}%
\providecommand \@@href[1]{\endgroup#1\@@endlink}%
\providecommand \@sanitize@url [0]{\catcode `\\12\catcode `\$12\catcode
  `\&12\catcode `\#12\catcode `\^12\catcode `\_12\catcode `\%12\relax}%
\providecommand \@@startlink[1]{}%
\providecommand \@@endlink[0]{}%
\providecommand \url  [0]{\begingroup\@sanitize@url \@url }%
\providecommand \@url [1]{\endgroup\@href {#1}{\urlprefix }}%
\providecommand \urlprefix  [0]{URL }%
\providecommand \Eprint [0]{\href }%
\providecommand \doibase [0]{https://doi.org/}%
\providecommand \selectlanguage [0]{\@gobble}%
\providecommand \bibinfo  [0]{\@secondoftwo}%
\providecommand \bibfield  [0]{\@secondoftwo}%
\providecommand \translation [1]{[#1]}%
\providecommand \BibitemOpen [0]{}%
\providecommand \bibitemStop [0]{}%
\providecommand \bibitemNoStop [0]{.\EOS\space}%
\providecommand \EOS [0]{\spacefactor3000\relax}%
\providecommand \BibitemShut  [1]{\csname bibitem#1\endcsname}%
\let\auto@bib@innerbib\@empty
\bibitem [{\citenamefont {Andrews}\ \emph {et~al.}(1997)\citenamefont
  {Andrews}, \citenamefont {Townsend}, \citenamefont {Miesner}, \citenamefont
  {Durfee}, \citenamefont {Kurn},\ and\ \citenamefont {Ketterle}}]{Andrews637}%
  \BibitemOpen
  \bibfield  {author} {\bibinfo {author} {\bibfnamefont {M.~R.}\ \bibnamefont
  {Andrews}}, \bibinfo {author} {\bibfnamefont {C.~G.}\ \bibnamefont
  {Townsend}}, \bibinfo {author} {\bibfnamefont {H.-J.}\ \bibnamefont
  {Miesner}}, \bibinfo {author} {\bibfnamefont {D.~S.}\ \bibnamefont {Durfee}},
  \bibinfo {author} {\bibfnamefont {D.~M.}\ \bibnamefont {Kurn}},\ and\
  \bibinfo {author} {\bibfnamefont {W.}~\bibnamefont {Ketterle}},\ }\bibfield
  {title} {\bibinfo {title} {Observation of interference between two Bose
  condensates},\ }\href
  {https://doi.org/https://doi.org/10.1126/science.275.5300.637} {\bibfield
  {journal} {\bibinfo  {journal} {Science}\ }\textbf {\bibinfo {volume}
  {275}},\ \bibinfo {pages} {637} (\bibinfo {year} {1997})}\BibitemShut
  {NoStop}%
\bibitem [{\citenamefont {Ketterle}(2002)}]{ketterle}%
  \BibitemOpen
  \bibfield  {author} {\bibinfo {author} {\bibfnamefont {W.}~\bibnamefont
  {Ketterle}},\ }\bibfield  {title} {\bibinfo {title} {Nobel lecture: When
  atoms behave as waves: Bose-Einstein condensation and the atom laser},\
  }\href {https://doi.org/10.1103/RevModPhys.74.1131} {\bibfield  {journal}
  {\bibinfo  {journal} {Rev. Mod. Phys.}\ }\textbf {\bibinfo {volume} {74}},\
  \bibinfo {pages} {1131} (\bibinfo {year} {2002})}\BibitemShut {NoStop}%
\bibitem [{\citenamefont {Javanainen}\ and\ \citenamefont
  {Wilkens}(1997)}]{javanainen1997phase}%
  \BibitemOpen
  \bibfield  {author} {\bibinfo {author} {\bibfnamefont {J.}~\bibnamefont
  {Javanainen}}\ and\ \bibinfo {author} {\bibfnamefont {M.}~\bibnamefont
  {Wilkens}},\ }\bibfield  {title} {\bibinfo {title} {Phase and phase diffusion
  of a split Bose-Einstein condensate},\ }\href
  {https://doi.org/10.1103/PhysRevLett.78.4675} {\bibfield  {journal} {\bibinfo
   {journal} {Phys. Rev. Lett.}\ }\textbf {\bibinfo {volume} {78}},\ \bibinfo
  {pages} {4675} (\bibinfo {year} {1997})}\BibitemShut {NoStop}%
\bibitem [{\citenamefont {Javanainen}\ and\ \citenamefont
  {Yoo}(1996)}]{javanainen1996quantum}%
  \BibitemOpen
  \bibfield  {author} {\bibinfo {author} {\bibfnamefont {J.}~\bibnamefont
  {Javanainen}}\ and\ \bibinfo {author} {\bibfnamefont {S.~M.}\ \bibnamefont
  {Yoo}},\ }\bibfield  {title} {\bibinfo {title} {Quantum phase of a
  Bose-Einstein condensate with an arbitrary number of atoms},\ }\href
  {https://doi.org/10.1103/PhysRevLett.76.161} {\bibfield  {journal} {\bibinfo
  {journal} {Phys. Rev. Lett.}\ }\textbf {\bibinfo {volume} {76}},\ \bibinfo
  {pages} {161} (\bibinfo {year} {1996})}\BibitemShut {NoStop}%
\bibitem [{\citenamefont {Dalfovo}\ \emph {et~al.}(1999)\citenamefont
  {Dalfovo}, \citenamefont {Giorgini}, \citenamefont {Pitaevskii},\ and\
  \citenamefont {Stringari}}]{dalfovo1999theory}%
  \BibitemOpen
  \bibfield  {author} {\bibinfo {author} {\bibfnamefont {F.}~\bibnamefont
  {Dalfovo}}, \bibinfo {author} {\bibfnamefont {S.}~\bibnamefont {Giorgini}},
  \bibinfo {author} {\bibfnamefont {L.~P.}\ \bibnamefont {Pitaevskii}},\ and\
  \bibinfo {author} {\bibfnamefont {S.}~\bibnamefont {Stringari}},\ }\bibfield
  {title} {\bibinfo {title} {Theory of Bose-Einstein condensation in trapped
  gases},\ }\href {https://doi.org/10.1103/RevModPhys.71.463} {\bibfield
  {journal} {\bibinfo  {journal} {Rev. Mod. Phys.}\ }\textbf {\bibinfo {volume}
  {71}},\ \bibinfo {pages} {463} (\bibinfo {year} {1999})}\BibitemShut
  {NoStop}%
\bibitem [{\citenamefont {Pitaevskii}\ and\ \citenamefont
  {Stringari}(2016)}]{pitaevskii2016bose}%
  \BibitemOpen
  \bibfield  {author} {\bibinfo {author} {\bibfnamefont {L.~P.}\ \bibnamefont
  {Pitaevskii}}\ and\ \bibinfo {author} {\bibfnamefont {S.}~\bibnamefont
  {Stringari}},\ }\href@noop {} {\emph {\bibinfo {title} {Bose-Einstein
  condensation and superfluidity}}},\ Vol.\ \bibinfo {volume} {164}\ (\bibinfo
  {publisher} {Oxford University Press},\ \bibinfo {year} {2016})\BibitemShut
  {NoStop}%
\bibitem [{\citenamefont {Pethick}\ and\ \citenamefont
  {Smith}(2008)}]{pethick2008bose}%
  \BibitemOpen
  \bibfield  {author} {\bibinfo {author} {\bibfnamefont {C.~J.}\ \bibnamefont
  {Pethick}}\ and\ \bibinfo {author} {\bibfnamefont {H.}~\bibnamefont
  {Smith}},\ }\href@noop {} {\emph {\bibinfo {title} {Bose--Einstein
  condensation in dilute gases}}}\ (\bibinfo  {publisher} {Cambridge university
  press},\ \bibinfo {year} {2008})\BibitemShut {NoStop}%
\bibitem [{\citenamefont {Fattori}\ \emph {et~al.}(2008)\citenamefont
  {Fattori}, \citenamefont {D'Errico}, \citenamefont {Roati}, \citenamefont
  {Zaccanti}, \citenamefont {Jona-Lasinio}, \citenamefont {Modugno},
  \citenamefont {Inguscio},\ and\ \citenamefont
  {Modugno}}]{PhysRevLett.100.080405}%
  \BibitemOpen
  \bibfield  {author} {\bibinfo {author} {\bibfnamefont {M.}~\bibnamefont
  {Fattori}}, \bibinfo {author} {\bibfnamefont {C.}~\bibnamefont {D'Errico}},
  \bibinfo {author} {\bibfnamefont {G.}~\bibnamefont {Roati}}, \bibinfo
  {author} {\bibfnamefont {M.}~\bibnamefont {Zaccanti}}, \bibinfo {author}
  {\bibfnamefont {M.}~\bibnamefont {Jona-Lasinio}}, \bibinfo {author}
  {\bibfnamefont {M.}~\bibnamefont {Modugno}}, \bibinfo {author} {\bibfnamefont
  {M.}~\bibnamefont {Inguscio}},\ and\ \bibinfo {author} {\bibfnamefont
  {G.}~\bibnamefont {Modugno}},\ }\bibfield  {title} {\bibinfo {title} {Atom
  interferometry with a weakly interacting Bose-Einstein condensate},\ }\href
  {https://doi.org/10.1103/PhysRevLett.100.080405} {\bibfield  {journal}
  {\bibinfo  {journal} {Phys. Rev. Lett.}\ }\textbf {\bibinfo {volume} {100}},\
  \bibinfo {pages} {080405} (\bibinfo {year} {2008})}\BibitemShut {NoStop}%
\bibitem [{\citenamefont {Mewes}\ \emph {et~al.}(1997)\citenamefont {Mewes},
  \citenamefont {Andrews}, \citenamefont {Kurn}, \citenamefont {Durfee},
  \citenamefont {Townsend},\ and\ \citenamefont {Ketterle}}]{mewes}%
  \BibitemOpen
  \bibfield  {author} {\bibinfo {author} {\bibfnamefont {M.-O.}\ \bibnamefont
  {Mewes}}, \bibinfo {author} {\bibfnamefont {M.~R.}\ \bibnamefont {Andrews}},
  \bibinfo {author} {\bibfnamefont {D.~M.}\ \bibnamefont {Kurn}}, \bibinfo
  {author} {\bibfnamefont {D.~S.}\ \bibnamefont {Durfee}}, \bibinfo {author}
  {\bibfnamefont {C.~G.}\ \bibnamefont {Townsend}},\ and\ \bibinfo {author}
  {\bibfnamefont {W.}~\bibnamefont {Ketterle}},\ }\bibfield  {title} {\bibinfo
  {title} {Output coupler for Bose-Einstein condensed atoms},\ }\href
  {https://doi.org/10.1103/PhysRevLett.78.582} {\bibfield  {journal} {\bibinfo
  {journal} {Phys. Rev. Lett.}\ }\textbf {\bibinfo {volume} {78}},\ \bibinfo
  {pages} {582} (\bibinfo {year} {1997})}\BibitemShut {NoStop}%
\bibitem [{\citenamefont {Lee}\ \emph {et~al.}(2012)\citenamefont {Lee},
  \citenamefont {Huang}, \citenamefont {Deng}, \citenamefont {Dai},\ and\
  \citenamefont {Xu}}]{Lee2012}%
  \BibitemOpen
  \bibfield  {author} {\bibinfo {author} {\bibfnamefont {C.}~\bibnamefont
  {Lee}}, \bibinfo {author} {\bibfnamefont {J.}~\bibnamefont {Huang}}, \bibinfo
  {author} {\bibfnamefont {H.}~\bibnamefont {Deng}}, \bibinfo {author}
  {\bibfnamefont {H.}~\bibnamefont {Dai}},\ and\ \bibinfo {author}
  {\bibfnamefont {J.}~\bibnamefont {Xu}},\ }\bibfield  {title} {\bibinfo
  {title} {Nonlinear quantum interferometry with Bose condensed atoms},\ }\href
  {https://doi.org/https://doi.org/10.1007/s11467-011-0228-6} {\bibfield
  {journal} {\bibinfo  {journal} {Front. Phys.}\ }\textbf {\bibinfo {volume}
  {7}},\ \bibinfo {pages} {109} (\bibinfo {year} {2012})}\BibitemShut {NoStop}%
\bibitem [{\citenamefont {Morsch}\ \emph {et~al.}(2002)\citenamefont {Morsch},
  \citenamefont {Cristiani}, \citenamefont {M\"uller}, \citenamefont
  {Ciampini},\ and\ \citenamefont {Arimondo}}]{PhysRevA.66.021601}%
  \BibitemOpen
  \bibfield  {author} {\bibinfo {author} {\bibfnamefont {O.}~\bibnamefont
  {Morsch}}, \bibinfo {author} {\bibfnamefont {M.}~\bibnamefont {Cristiani}},
  \bibinfo {author} {\bibfnamefont {J.~H.}\ \bibnamefont {M\"uller}}, \bibinfo
  {author} {\bibfnamefont {D.}~\bibnamefont {Ciampini}},\ and\ \bibinfo
  {author} {\bibfnamefont {E.}~\bibnamefont {Arimondo}},\ }\bibfield  {title}
  {\bibinfo {title} {Free expansion of a Bose-Einstein condensate in a
  one-dimensional optical lattice},\ }\href
  {https://doi.org/10.1103/PhysRevA.66.021601} {\bibfield  {journal} {\bibinfo
  {journal} {Phys. Rev. A}\ }\textbf {\bibinfo {volume} {66}},\ \bibinfo
  {pages} {021601(R)} (\bibinfo {year} {2002})}\BibitemShut {NoStop}%
\bibitem [{\citenamefont {Morsch}\ and\ \citenamefont
  {Arimondo}(2002)}]{Morsch2002}%
  \BibitemOpen
  \bibfield  {author} {\bibinfo {author} {\bibfnamefont {O.}~\bibnamefont
  {Morsch}}\ and\ \bibinfo {author} {\bibfnamefont {E.}~\bibnamefont
  {Arimondo}},\ }\bibinfo {title} {Ultracold atoms and Bose-Einstein
  condensates in optical lattices},\ in\ \href@noop {} {\emph {\bibinfo
  {booktitle} {Dynamics and Thermodynamics of Systems with Long-Range
  Interactions}}},\ \bibinfo {editor} {edited by\ \bibinfo {editor}
  {\bibfnamefont {T.}~\bibnamefont {Dauxois}}, \bibinfo {editor} {\bibfnamefont
  {S.}~\bibnamefont {Ruffo}}, \bibinfo {editor} {\bibfnamefont
  {E.}~\bibnamefont {Arimondo}},\ and\ \bibinfo {editor} {\bibfnamefont
  {M.}~\bibnamefont {Wilkens}}}\ (\bibinfo  {publisher} {Springer Berlin
  Heidelberg},\ \bibinfo {address} {Berlin, Heidelberg},\ \bibinfo {year}
  {2002})\ pp.\ \bibinfo {pages} {312--331}\BibitemShut {NoStop}%
\bibitem [{\citenamefont {Bao}\ \emph {et~al.}(2003)\citenamefont {Bao},
  \citenamefont {Jaksch},\ and\ \citenamefont {Markowich}}]{bao2003numerical}%
  \BibitemOpen
  \bibfield  {author} {\bibinfo {author} {\bibfnamefont {W.}~\bibnamefont
  {Bao}}, \bibinfo {author} {\bibfnamefont {D.}~\bibnamefont {Jaksch}},\ and\
  \bibinfo {author} {\bibfnamefont {P.~A.}\ \bibnamefont {Markowich}},\
  }\bibfield  {title} {\bibinfo {title} {Numerical solution of the
  Gross-Pitaevskii equation for Bose-Einstein condensation},\ }\href
  {https://doi.org/https://doi.org/10.1016/S0021-9991(03)00102-5} {\bibfield
  {journal} {\bibinfo  {journal} {J. Comput. Phys.}\ }\textbf {\bibinfo
  {volume} {187}},\ \bibinfo {pages} {318} (\bibinfo {year}
  {2003})}\BibitemShut {NoStop}%
\bibitem [{\citenamefont {Antoine}\ \emph {et~al.}(2013)\citenamefont
  {Antoine}, \citenamefont {Bao},\ and\ \citenamefont
  {Besse}}]{ANTOINE20132621}%
  \BibitemOpen
  \bibfield  {author} {\bibinfo {author} {\bibfnamefont {X.}~\bibnamefont
  {Antoine}}, \bibinfo {author} {\bibfnamefont {W.}~\bibnamefont {Bao}},\ and\
  \bibinfo {author} {\bibfnamefont {C.}~\bibnamefont {Besse}},\ }\bibfield
  {title} {\bibinfo {title} {Computational methods for the dynamics of the
  nonlinear Schr\"odinger/Gross-Pitaevskii equations},\ }\href
  {https://doi.org/https://doi.org/10.1016/j.cpc.2013.07.012} {\bibfield
  {journal} {\bibinfo  {journal} {Comp. Phys. Comm.}\ }\textbf {\bibinfo
  {volume} {184}},\ \bibinfo {pages} {2621} (\bibinfo {year}
  {2013})}\BibitemShut {NoStop}%
\bibitem [{\citenamefont {Cerimele}\ \emph {et~al.}(2000)\citenamefont
  {Cerimele}, \citenamefont {Chiofalo}, \citenamefont {Pistella}, \citenamefont
  {Succi},\ and\ \citenamefont {Tosi}}]{PhysRevE.62.1382}%
  \BibitemOpen
  \bibfield  {author} {\bibinfo {author} {\bibfnamefont {M.~M.}\ \bibnamefont
  {Cerimele}}, \bibinfo {author} {\bibfnamefont {M.~L.}\ \bibnamefont
  {Chiofalo}}, \bibinfo {author} {\bibfnamefont {F.}~\bibnamefont {Pistella}},
  \bibinfo {author} {\bibfnamefont {S.}~\bibnamefont {Succi}},\ and\ \bibinfo
  {author} {\bibfnamefont {M.~P.}\ \bibnamefont {Tosi}},\ }\bibfield  {title}
  {\bibinfo {title} {Numerical solution of the Gross-Pitaevskii equation using
  an explicit finite-difference scheme: An application to trapped Bose-Einstein
  condensates},\ }\href {https://doi.org/10.1103/PhysRevE.62.1382} {\bibfield
  {journal} {\bibinfo  {journal} {Phys. Rev. E}\ }\textbf {\bibinfo {volume}
  {62}},\ \bibinfo {pages} {1382} (\bibinfo {year} {2000})}\BibitemShut
  {NoStop}%
\bibitem [{\citenamefont {Polo}\ and\ \citenamefont
  {Ahufinger}(2013)}]{polo2013soliton}%
  \BibitemOpen
  \bibfield  {author} {\bibinfo {author} {\bibfnamefont {J.}~\bibnamefont
  {Polo}}\ and\ \bibinfo {author} {\bibfnamefont {V.}~\bibnamefont
  {Ahufinger}},\ }\bibfield  {title} {\bibinfo {title} {Soliton-based
  matter-wave interferometer},\ }\href
  {https://doi.org/10.1103/PhysRevA.88.053628} {\bibfield  {journal} {\bibinfo
  {journal} {Phys. Rev. A}\ }\textbf {\bibinfo {volume} {88}},\ \bibinfo
  {pages} {053628} (\bibinfo {year} {2013})}\BibitemShut {NoStop}%
\bibitem [{\citenamefont {Ji}\ \emph {et~al.}(2016)\citenamefont {Ji},
  \citenamefont {Wang}, \citenamefont {Luo},\ and\ \citenamefont
  {Liu}}]{Ji_2016}%
  \BibitemOpen
  \bibfield  {author} {\bibinfo {author} {\bibfnamefont {S.-T.}\ \bibnamefont
  {Ji}}, \bibinfo {author} {\bibfnamefont {Y.-S.}\ \bibnamefont {Wang}},
  \bibinfo {author} {\bibfnamefont {Y.-E.}\ \bibnamefont {Luo}},\ and\ \bibinfo
  {author} {\bibfnamefont {X.-S.}\ \bibnamefont {Liu}},\ }\bibfield  {title}
  {\bibinfo {title} {Generating periodic interference in Bose-Einstein
  condensates},\ }\href {https://doi.org/10.1088/1674-1056/25/9/090303}
  {\bibfield  {journal} {\bibinfo  {journal} {Chin. Phys. B}\ }\textbf
  {\bibinfo {volume} {25}},\ \bibinfo {pages} {090303} (\bibinfo {year}
  {2016})}\BibitemShut {NoStop}%
\bibitem [{\citenamefont {Ahufinger}\ \emph {et~al.}(2004)\citenamefont
  {Ahufinger}, \citenamefont {Sanpera}, \citenamefont {Pedri}, \citenamefont
  {Santos},\ and\ \citenamefont {Lewenstein}}]{ahufinger1}%
  \BibitemOpen
  \bibfield  {author} {\bibinfo {author} {\bibfnamefont {V.}~\bibnamefont
  {Ahufinger}}, \bibinfo {author} {\bibfnamefont {A.}~\bibnamefont {Sanpera}},
  \bibinfo {author} {\bibfnamefont {P.}~\bibnamefont {Pedri}}, \bibinfo
  {author} {\bibfnamefont {L.}~\bibnamefont {Santos}},\ and\ \bibinfo {author}
  {\bibfnamefont {M.}~\bibnamefont {Lewenstein}},\ }\bibfield  {title}
  {\bibinfo {title} {Creation and mobility of discrete solitons in
  Bose-Einstein condensates},\ }\href
  {https://doi.org/10.1103/PhysRevA.69.053604} {\bibfield  {journal} {\bibinfo
  {journal} {Phys. Rev. A}\ }\textbf {\bibinfo {volume} {69}},\ \bibinfo
  {pages} {053604} (\bibinfo {year} {2004})}\BibitemShut {NoStop}%
\bibitem [{\citenamefont {Ottaviani}\ \emph {et~al.}(2010)\citenamefont
  {Ottaviani}, \citenamefont {Ahufinger}, \citenamefont {Corbal\'an},\ and\
  \citenamefont {Mompart}}]{ahufinger2}%
  \BibitemOpen
  \bibfield  {author} {\bibinfo {author} {\bibfnamefont {C.}~\bibnamefont
  {Ottaviani}}, \bibinfo {author} {\bibfnamefont {V.}~\bibnamefont
  {Ahufinger}}, \bibinfo {author} {\bibfnamefont {R.}~\bibnamefont
  {Corbal\'an}},\ and\ \bibinfo {author} {\bibfnamefont {J.}~\bibnamefont
  {Mompart}},\ }\bibfield  {title} {\bibinfo {title} {Adiabatic splitting,
  transport, and self-trapping of a Bose-Einstein condensate in a double-well
  potential},\ }\href {https://doi.org/10.1103/PhysRevA.81.043621} {\bibfield
  {journal} {\bibinfo  {journal} {Phys. Rev. A}\ }\textbf {\bibinfo {volume}
  {81}},\ \bibinfo {pages} {043621} (\bibinfo {year} {2010})}\BibitemShut
  {NoStop}%
\bibitem [{\citenamefont {Berry}\ and\ \citenamefont
  {Kutz}(2007)}]{PhysRevE.75.036214}%
  \BibitemOpen
  \bibfield  {author} {\bibinfo {author} {\bibfnamefont {N.~H.}\ \bibnamefont
  {Berry}}\ and\ \bibinfo {author} {\bibfnamefont {J.~N.}\ \bibnamefont
  {Kutz}},\ }\bibfield  {title} {\bibinfo {title} {Dynamics of Bose-Einstein
  condensates under the influence of periodic and harmonic potentials},\ }\href
  {https://doi.org/10.1103/PhysRevE.75.036214} {\bibfield  {journal} {\bibinfo
  {journal} {Phys. Rev. E}\ }\textbf {\bibinfo {volume} {75}},\ \bibinfo
  {pages} {036214} (\bibinfo {year} {2007})}\BibitemShut {NoStop}%
\bibitem [{\citenamefont {Yan}\ and\ \citenamefont
  {Jiang}(2012)}]{PhysRevE.85.056608}%
  \BibitemOpen
  \bibfield  {author} {\bibinfo {author} {\bibfnamefont {Z.}~\bibnamefont
  {Yan}}\ and\ \bibinfo {author} {\bibfnamefont {D.}~\bibnamefont {Jiang}},\
  }\bibfield  {title} {\bibinfo {title} {Matter-wave solutions in Bose-Einstein
  condensates with harmonic and Gaussian potentials},\ }\href
  {https://doi.org/10.1103/PhysRevE.85.056608} {\bibfield  {journal} {\bibinfo
  {journal} {Phys. Rev. E}\ }\textbf {\bibinfo {volume} {85}},\ \bibinfo
  {pages} {056608} (\bibinfo {year} {2012})}\BibitemShut {NoStop}%
\bibitem [{\citenamefont {Mallory}\ and\ \citenamefont
  {Van~Gorder}(2014)}]{PhysRevE.89.013204}%
  \BibitemOpen
  \bibfield  {author} {\bibinfo {author} {\bibfnamefont {K.}~\bibnamefont
  {Mallory}}\ and\ \bibinfo {author} {\bibfnamefont {R.~A.}\ \bibnamefont
  {Van~Gorder}},\ }\bibfield  {title} {\bibinfo {title} {Stationary solutions
  for the 1$+$1 nonlinear schr\"odinger equation modeling attractive
  Bose-Einstein condensates in small potentials},\ }\href
  {https://doi.org/10.1103/PhysRevE.89.013204} {\bibfield  {journal} {\bibinfo
  {journal} {Phys. Rev. E}\ }\textbf {\bibinfo {volume} {89}},\ \bibinfo
  {pages} {013204} (\bibinfo {year} {2014})}\BibitemShut {NoStop}%
\bibitem [{\citenamefont {Sanz}\ and\ \citenamefont
  {Miret-Art\'es}(2012)}]{sanz-bk-1}%
  \BibitemOpen
  \bibfield  {author} {\bibinfo {author} {\bibfnamefont {A.~S.}\ \bibnamefont
  {Sanz}}\ and\ \bibinfo {author} {\bibfnamefont {S.}~\bibnamefont
  {Miret-Art\'es}},\ }\href@noop {} {\emph {\bibinfo {title} {A Trajectory
  Description of Quantum Processes. I. Fundamentals}}},\ \bibinfo {series}
  {Lecture Notes in Physics}, Vol.\ \bibinfo {volume} {850}\ (\bibinfo
  {publisher} {Springer},\ \bibinfo {address} {Berlin},\ \bibinfo {year}
  {2012})\BibitemShut {NoStop}%
\bibitem [{\citenamefont {Sanz}(2019)}]{sanz:FrontPhys:2019}%
  \BibitemOpen
  \bibfield  {author} {\bibinfo {author} {\bibfnamefont {A.~S.}\ \bibnamefont
  {Sanz}},\ }\bibfield  {title} {\bibinfo {title} {Bohm's approach to quantum
  mechanics: Alternative theory or practical picture?},\ }\href
  {https://doi.org/https://doi.org/10.1007/s11467-018-0853-4} {\bibfield
  {journal} {\bibinfo  {journal} {Front. Phys.}\ }\textbf {\bibinfo {volume}
  {14}},\ \bibinfo {pages} {11301} (\bibinfo {year} {2019})}\BibitemShut
  {NoStop}%
\bibitem [{\citenamefont {Benseny}\ \emph {et~al.}(2012)\citenamefont
  {Benseny}, \citenamefont {Bagud\`a}, \citenamefont {Oriols}, \citenamefont
  {Birkl},\ and\ \citenamefont {Mompart}}]{benseny:PanStanford:2012}%
  \BibitemOpen
  \bibfield  {author} {\bibinfo {author} {\bibfnamefont {A.}~\bibnamefont
  {Benseny}}, \bibinfo {author} {\bibfnamefont {J.}~\bibnamefont {Bagud\`a}},
  \bibinfo {author} {\bibfnamefont {X.}~\bibnamefont {Oriols}}, \bibinfo
  {author} {\bibfnamefont {G.}~\bibnamefont {Birkl}},\ and\ \bibinfo {author}
  {\bibfnamefont {J.}~\bibnamefont {Mompart}},\ }\href@noop {} {\emph {\bibinfo
  {title} {Applied Bohmian Mechanics: From Nanoscale Systems to Cosmology}}},\
  edited by\ \bibinfo {editor} {\bibfnamefont {X.}~\bibnamefont {Oriols}}\ and\
  \bibinfo {editor} {\bibfnamefont {J.}~\bibnamefont {Mompart}}\ (\bibinfo
  {publisher} {Pan Standford Publishing},\ \bibinfo {address} {Singapore},\
  \bibinfo {year} {2012})\ Chap.\ \bibinfo {chapter} {Atomtronics: Coherent
  control of atomic flow via adiabatic passage}, pp.\ \bibinfo {pages}
  {189--233}\BibitemShut {NoStop}%
\bibitem [{\citenamefont {Cornish}\ \emph {et~al.}(2000)\citenamefont
  {Cornish}, \citenamefont {Claussen}, \citenamefont {Roberts}, \citenamefont
  {Cornell},\ and\ \citenamefont {Wieman}}]{PhysRevLett.85.1795}%
  \BibitemOpen
  \bibfield  {author} {\bibinfo {author} {\bibfnamefont {S.~L.}\ \bibnamefont
  {Cornish}}, \bibinfo {author} {\bibfnamefont {N.~R.}\ \bibnamefont
  {Claussen}}, \bibinfo {author} {\bibfnamefont {J.~L.}\ \bibnamefont
  {Roberts}}, \bibinfo {author} {\bibfnamefont {E.~A.}\ \bibnamefont
  {Cornell}},\ and\ \bibinfo {author} {\bibfnamefont {C.~E.}\ \bibnamefont
  {Wieman}},\ }\bibfield  {title} {\bibinfo {title} {Stable
  ${}^{85}\mathrm{Rb}$ Bose-Einstein condensates with widely tunable
  interactions},\ }\href {https://doi.org/10.1103/PhysRevLett.85.1795}
  {\bibfield  {journal} {\bibinfo  {journal} {Phys. Rev. Lett.}\ }\textbf
  {\bibinfo {volume} {85}},\ \bibinfo {pages} {1795} (\bibinfo {year}
  {2000})}\BibitemShut {NoStop}%
\bibitem [{\citenamefont {Bradley}\ \emph {et~al.}(1995)\citenamefont
  {Bradley}, \citenamefont {Sackett}, \citenamefont {Tollett},\ and\
  \citenamefont {Hulet}}]{PhysRevLett.75.1687}%
  \BibitemOpen
  \bibfield  {author} {\bibinfo {author} {\bibfnamefont {C.~C.}\ \bibnamefont
  {Bradley}}, \bibinfo {author} {\bibfnamefont {C.~A.}\ \bibnamefont
  {Sackett}}, \bibinfo {author} {\bibfnamefont {J.~J.}\ \bibnamefont
  {Tollett}},\ and\ \bibinfo {author} {\bibfnamefont {R.~G.}\ \bibnamefont
  {Hulet}},\ }\bibfield  {title} {\bibinfo {title} {Evidence of Bose-Einstein
  condensation in an atomic gas with attractive interactions},\ }\href
  {https://doi.org/10.1103/PhysRevLett.75.1687} {\bibfield  {journal} {\bibinfo
   {journal} {Phys. Rev. Lett.}\ }\textbf {\bibinfo {volume} {75}},\ \bibinfo
  {pages} {1687} (\bibinfo {year} {1995})}\BibitemShut {NoStop}%
\bibitem [{\citenamefont {Anderson}\ \emph {et~al.}(1995)\citenamefont
  {Anderson}, \citenamefont {Ensher}, \citenamefont {Matthews}, \citenamefont
  {Wieman},\ and\ \citenamefont {Cornell}}]{Anderson198}%
  \BibitemOpen
  \bibfield  {author} {\bibinfo {author} {\bibfnamefont {M.~H.}\ \bibnamefont
  {Anderson}}, \bibinfo {author} {\bibfnamefont {J.~R.}\ \bibnamefont
  {Ensher}}, \bibinfo {author} {\bibfnamefont {M.~R.}\ \bibnamefont
  {Matthews}}, \bibinfo {author} {\bibfnamefont {C.~E.}\ \bibnamefont
  {Wieman}},\ and\ \bibinfo {author} {\bibfnamefont {E.~A.}\ \bibnamefont
  {Cornell}},\ }\bibfield  {title} {\bibinfo {title} {Observation of
  Bose-Einstein condensation in a dilute atomic vapor},\ }\href
  {https://doi.org/10.1126/science.269.5221.198} {\bibfield  {journal}
  {\bibinfo  {journal} {Science}\ }\textbf {\bibinfo {volume} {269}},\ \bibinfo
  {pages} {198} (\bibinfo {year} {1995})}\BibitemShut {NoStop}%
\bibitem [{\citenamefont {Davis}\ \emph {et~al.}(1995)\citenamefont {Davis},
  \citenamefont {Mewes}, \citenamefont {Andrews}, \citenamefont {van Druten},
  \citenamefont {Durfee}, \citenamefont {Kurn},\ and\ \citenamefont
  {Ketterle}}]{PhysRevLett.75.3969}%
  \BibitemOpen
  \bibfield  {author} {\bibinfo {author} {\bibfnamefont {K.~B.}\ \bibnamefont
  {Davis}}, \bibinfo {author} {\bibfnamefont {M.~O.}\ \bibnamefont {Mewes}},
  \bibinfo {author} {\bibfnamefont {M.~R.}\ \bibnamefont {Andrews}}, \bibinfo
  {author} {\bibfnamefont {N.~J.}\ \bibnamefont {van Druten}}, \bibinfo
  {author} {\bibfnamefont {D.~S.}\ \bibnamefont {Durfee}}, \bibinfo {author}
  {\bibfnamefont {D.~M.}\ \bibnamefont {Kurn}},\ and\ \bibinfo {author}
  {\bibfnamefont {W.}~\bibnamefont {Ketterle}},\ }\bibfield  {title} {\bibinfo
  {title} {Bose-Einstein condensation in a gas of sodium atoms},\ }\href
  {https://doi.org/10.1103/PhysRevLett.75.3969} {\bibfield  {journal} {\bibinfo
   {journal} {Phys. Rev. Lett.}\ }\textbf {\bibinfo {volume} {75}},\ \bibinfo
  {pages} {3969} (\bibinfo {year} {1995})}\BibitemShut {NoStop}%
\bibitem [{\citenamefont {Burger}\ \emph {et~al.}(1999)\citenamefont {Burger},
  \citenamefont {Bongs}, \citenamefont {Dettmer}, \citenamefont {Ertmer},
  \citenamefont {Sengstock}, \citenamefont {Sanpera}, \citenamefont
  {Shlyapnikov},\ and\ \citenamefont {Lewenstein}}]{lewenstein:PRL:1999}%
  \BibitemOpen
  \bibfield  {author} {\bibinfo {author} {\bibfnamefont {S.}~\bibnamefont
  {Burger}}, \bibinfo {author} {\bibfnamefont {K.}~\bibnamefont {Bongs}},
  \bibinfo {author} {\bibfnamefont {S.}~\bibnamefont {Dettmer}}, \bibinfo
  {author} {\bibfnamefont {W.}~\bibnamefont {Ertmer}}, \bibinfo {author}
  {\bibfnamefont {K.}~\bibnamefont {Sengstock}}, \bibinfo {author}
  {\bibfnamefont {A.}~\bibnamefont {Sanpera}}, \bibinfo {author} {\bibfnamefont
  {G.~V.}\ \bibnamefont {Shlyapnikov}},\ and\ \bibinfo {author} {\bibfnamefont
  {M.}~\bibnamefont {Lewenstein}},\ }\bibfield  {title} {\bibinfo {title} {Dark
  solitons in Bose-Einstein condensates},\ }\href
  {https://doi.org/10.1103/PhysRevLett.83.5198} {\bibfield  {journal} {\bibinfo
   {journal} {Phys. Rev. Lett.}\ }\textbf {\bibinfo {volume} {83}},\ \bibinfo
  {pages} {5198} (\bibinfo {year} {1999})}\BibitemShut {NoStop}%
\bibitem [{\citenamefont {Scott}\ \emph {et~al.}(1998)\citenamefont {Scott},
  \citenamefont {Ballagh},\ and\ \citenamefont {Burnett}}]{burnett:JPB:1998}%
  \BibitemOpen
  \bibfield  {author} {\bibinfo {author} {\bibfnamefont {T.~F.}\ \bibnamefont
  {Scott}}, \bibinfo {author} {\bibfnamefont {R.~J.}\ \bibnamefont {Ballagh}},\
  and\ \bibinfo {author} {\bibfnamefont {K.}~\bibnamefont {Burnett}},\
  }\bibfield  {title} {\bibinfo {title} {Formation of fundamental structures in
  bose--einstein condensates},\ }\href
  {https://doi.org/10.1088/0953-4075/31/8/001} {\bibfield  {journal} {\bibinfo
  {journal} {J. Phys. B: At. Mol. Opt. Phys.}\ }\textbf {\bibinfo {volume}
  {31}},\ \bibinfo {pages} {L329} (\bibinfo {year} {1998})}\BibitemShut
  {NoStop}%
\bibitem [{\citenamefont {Dobrek}\ \emph {et~al.}(1999)\citenamefont {Dobrek},
  \citenamefont {Gajda}, \citenamefont {Lewenstein}, \citenamefont {Sengstock},
  \citenamefont {Birkl},\ and\ \citenamefont {Ertmer}}]{dobrek:PRA:1999}%
  \BibitemOpen
  \bibfield  {author} {\bibinfo {author} {\bibfnamefont {{\L}.}~\bibnamefont
  {Dobrek}}, \bibinfo {author} {\bibfnamefont {M.}~\bibnamefont {Gajda}},
  \bibinfo {author} {\bibfnamefont {M.}~\bibnamefont {Lewenstein}}, \bibinfo
  {author} {\bibfnamefont {K.}~\bibnamefont {Sengstock}}, \bibinfo {author}
  {\bibfnamefont {G.}~\bibnamefont {Birkl}},\ and\ \bibinfo {author}
  {\bibfnamefont {W.}~\bibnamefont {Ertmer}},\ }\bibfield  {title} {\bibinfo
  {title} {Optical generation of vortices in trapped Bose-Einstein
  condensates},\ }\href {https://doi.org/10.1103/PhysRevA.60.R3381} {\bibfield
  {journal} {\bibinfo  {journal} {Phys. Rev. A}\ }\textbf {\bibinfo {volume}
  {60}},\ \bibinfo {pages} {R3381} (\bibinfo {year} {1999})}\BibitemShut
  {NoStop}%
\bibitem [{\citenamefont {Kevrekidis}\ \emph {et~al.}(2008)\citenamefont
  {Kevrekidis}, \citenamefont {Frantzeskakis},\ and\ \citenamefont
  {Carretero-Gonz{\'a}lez}}]{Frantzeskakis-bk:2008}%
  \BibitemOpen
  \bibinfo {editor} {\bibfnamefont {P.~G.}\ \bibnamefont {Kevrekidis}},
  \bibinfo {editor} {\bibfnamefont {D.~J.}\ \bibnamefont {Frantzeskakis}},\
  and\ \bibinfo {editor} {\bibfnamefont {R.}~\bibnamefont
  {Carretero-Gonz{\'a}lez}},\ eds.,\ \href
  {https://doi.org/10.1007/978-3-540-73591-5_4} {\emph {\bibinfo {title}
  {Emergent Nonlinear Phenomena in Bose-Einstein Condensates: Theory and
  Experiment}}}\ (\bibinfo  {publisher} {Springer Berlin Heidelberg},\ \bibinfo
  {address} {Berlin, Heidelberg},\ \bibinfo {year} {2008})\BibitemShut
  {NoStop}%
\bibitem [{\citenamefont {Frantzeskakis}(2010)}]{Frantzeskakis:JPA:2010}%
  \BibitemOpen
  \bibfield  {author} {\bibinfo {author} {\bibfnamefont {D.~J.}\ \bibnamefont
  {Frantzeskakis}},\ }\bibfield  {title} {\bibinfo {title} {Dark solitons in
  atomic Bose–Einstein condensates: from theory to experiments},\ }\href
  {https://doi.org/10.1088/1751-8113/43/21/213001} {\bibfield  {journal}
  {\bibinfo  {journal} {J. Phys. A: Math. Theor.}\ }\textbf {\bibinfo {volume}
  {43}},\ \bibinfo {pages} {213001} (\bibinfo {year} {2010})}\BibitemShut
  {NoStop}%
\bibitem [{\citenamefont {Tsuzuki}(1971)}]{tsuzuki:JLTP:1971}%
  \BibitemOpen
  \bibfield  {author} {\bibinfo {author} {\bibfnamefont {T.}~\bibnamefont
  {Tsuzuki}},\ }\bibfield  {title} {\bibinfo {title} {Nonlinear waves in the
  pitaevskii-gross equation},\ }\href {https://doi.org/10.1007/BF00628744}
  {\bibfield  {journal} {\bibinfo  {journal} {J. Low Temp. Phys.}\ }\textbf
  {\bibinfo {volume} {4}},\ \bibinfo {pages} {441} (\bibinfo {year}
  {1971})}\BibitemShut {NoStop}%
\bibitem [{\citenamefont {Parker}\ \emph {et~al.}(2003)\citenamefont {Parker},
  \citenamefont {Proukakis}, \citenamefont {Leadbeater},\ and\ \citenamefont
  {Adams}}]{parker2003soliton}%
  \BibitemOpen
  \bibfield  {author} {\bibinfo {author} {\bibfnamefont {N.~G.}\ \bibnamefont
  {Parker}}, \bibinfo {author} {\bibfnamefont {N.~P.}\ \bibnamefont
  {Proukakis}}, \bibinfo {author} {\bibfnamefont {M.}~\bibnamefont
  {Leadbeater}},\ and\ \bibinfo {author} {\bibfnamefont {C.~S.}\ \bibnamefont
  {Adams}},\ }\bibfield  {title} {\bibinfo {title} {Soliton-sound interactions
  in quasi-one-dimensional Bose-Einstein condensates},\ }\href
  {https://doi.org/10.1103/PhysRevLett.90.220401} {\bibfield  {journal}
  {\bibinfo  {journal} {Phys. Rev. Lett.}\ }\textbf {\bibinfo {volume} {90}},\
  \bibinfo {pages} {220401} (\bibinfo {year} {2003})}\BibitemShut {NoStop}%
\bibitem [{\citenamefont {Ronzheimer}(2008)}]{ronzheimer:PhDThesis:2008}%
  \BibitemOpen
  \bibfield  {author} {\bibinfo {author} {\bibfnamefont {J.~P.}\ \bibnamefont
  {Ronzheimer}},\ }\emph {\bibinfo {title} {Interactions of Dark Solitons in
  Cigar Shaped Bose-Einstein Condensates}},\ \href@noop {} {Ph.D. thesis},\
  \bibinfo  {school} {Kirchhoff-Institut f\"ur Physik, University of
  Heidelberg}, \bibinfo {address} {Germany} (\bibinfo {year}
  {2008})\BibitemShut {NoStop}%
\bibitem [{\citenamefont {Parker}\ \emph {et~al.}(2010)\citenamefont {Parker},
  \citenamefont {Proukakis},\ and\ \citenamefont {Adams}}]{parker2010dark}%
  \BibitemOpen
  \bibfield  {author} {\bibinfo {author} {\bibfnamefont {N.~G.}\ \bibnamefont
  {Parker}}, \bibinfo {author} {\bibfnamefont {N.~P.}\ \bibnamefont
  {Proukakis}},\ and\ \bibinfo {author} {\bibfnamefont {C.~S.}\ \bibnamefont
  {Adams}},\ }\bibfield  {title} {\bibinfo {title} {Dark soliton decay due to
  trap anharmonicity in atomic Bose-Einstein condensates},\ }\href
  {https://doi.org/10.1103/PhysRevA.81.033606} {\bibfield  {journal} {\bibinfo
  {journal} {Phys. Rev. A}\ }\textbf {\bibinfo {volume} {81}},\ \bibinfo
  {pages} {033606} (\bibinfo {year} {2010})}\BibitemShut {NoStop}%
\bibitem [{\citenamefont {Madelung}(1926)}]{madelung:ZPhys:1926}%
  \BibitemOpen
  \bibfield  {author} {\bibinfo {author} {\bibfnamefont {E.}~\bibnamefont
  {Madelung}},\ }\bibfield  {title} {\bibinfo {title} {Quantentheorie in
  hydrodynamischer form},\ }\href
  {https://doi.org/https://doi.org/10.1007/BF01400372} {\bibfield  {journal}
  {\bibinfo  {journal} {Z. Phys.}\ }\textbf {\bibinfo {volume} {40}},\ \bibinfo
  {pages} {322} (\bibinfo {year} {1926})}\BibitemShut {NoStop}%
\bibitem [{\citenamefont {Bohm}(1952{\natexlab{a}})}]{bohm:PR:1952-1}%
  \BibitemOpen
  \bibfield  {author} {\bibinfo {author} {\bibfnamefont {D.}~\bibnamefont
  {Bohm}},\ }\bibfield  {title} {\bibinfo {title} {A suggested interpretation
  of the quantum theory in terms of ``hidden'' variables. I},\ }\href
  {https://doi.org/10.1103/PhysRev.85.166} {\bibfield  {journal} {\bibinfo
  {journal} {Phys. Rev.}\ }\textbf {\bibinfo {volume} {85}},\ \bibinfo {pages}
  {166} (\bibinfo {year} {1952}{\natexlab{a}})}\BibitemShut {NoStop}%
\bibitem [{\citenamefont {Bohm}(1952{\natexlab{b}})}]{bohm:PR:1952-2}%
  \BibitemOpen
  \bibfield  {author} {\bibinfo {author} {\bibfnamefont {D.}~\bibnamefont
  {Bohm}},\ }\bibfield  {title} {\bibinfo {title} {A suggested interpretation
  of the quantum theory in terms of ``hidden'' variables. II},\ }\href
  {https://doi.org/10.1103/PhysRev.85.180} {\bibfield  {journal} {\bibinfo
  {journal} {Phys. Rev.}\ }\textbf {\bibinfo {volume} {85}},\ \bibinfo {pages}
  {180} (\bibinfo {year} {1952}{\natexlab{b}})}\BibitemShut {NoStop}%
\bibitem [{\citenamefont {Schiff}(1968)}]{schiff-bk}%
  \BibitemOpen
  \bibfield  {author} {\bibinfo {author} {\bibfnamefont {L.~I.}\ \bibnamefont
  {Schiff}},\ }\href@noop {} {\emph {\bibinfo {title} {Quantum Mechanics}}},\
  \bibinfo {edition} {3rd}\ ed.\ (\bibinfo  {publisher} {McGraw-Hill},\
  \bibinfo {address} {Singapore},\ \bibinfo {year} {1968})\BibitemShut
  {NoStop}%
\bibitem [{\citenamefont {Holland}(1993)}]{holland-bk}%
  \BibitemOpen
  \bibfield  {author} {\bibinfo {author} {\bibfnamefont {P.~R.}\ \bibnamefont
  {Holland}},\ }\href@noop {} {\emph {\bibinfo {title} {The Quantum Theory of
  Motion}}}\ (\bibinfo  {publisher} {Cambridge University Press},\ \bibinfo
  {address} {Cambridge},\ \bibinfo {year} {1993})\BibitemShut {NoStop}%
\bibitem [{\citenamefont {Landau}(1941{\natexlab{a}})}]{landau:JPUSSR:1941}%
  \BibitemOpen
  \bibfield  {author} {\bibinfo {author} {\bibfnamefont {L.~D.}\ \bibnamefont
  {Landau}},\ }\bibfield  {title} {\bibinfo {title} {The theory of
  superfluidity of helium II},\ }\href@noop {} {\bibfield  {journal} {\bibinfo
  {journal} {J. Phys. (USSR)}\ }\textbf {\bibinfo {volume} {5}},\ \bibinfo
  {pages} {71} (\bibinfo {year} {1941}{\natexlab{a}})}\BibitemShut {NoStop}%
\bibitem [{\citenamefont {Landau}(1941{\natexlab{b}})}]{landau:PhysRev:1941}%
  \BibitemOpen
  \bibfield  {author} {\bibinfo {author} {\bibfnamefont {L.}~\bibnamefont
  {Landau}},\ }\bibfield  {title} {\bibinfo {title} {Theory of the
  superfluidity of helium II},\ }\href {https://doi.org/10.1103/PhysRev.60.356}
  {\bibfield  {journal} {\bibinfo  {journal} {Phys. Rev.}\ }\textbf {\bibinfo
  {volume} {60}},\ \bibinfo {pages} {356} (\bibinfo {year}
  {1941}{\natexlab{b}})}\BibitemShut {NoStop}%
\bibitem [{\citenamefont {Lewenstein}\ \emph {et~al.}(2007)\citenamefont
  {Lewenstein}, \citenamefont {Sanpera}, \citenamefont {Ahufinger},
  \citenamefont {Damski}, \citenamefont {Sen(De)},\ and\ \citenamefont
  {Sen}}]{lewenstein:2007:AdvPhys}%
  \BibitemOpen
  \bibfield  {author} {\bibinfo {author} {\bibfnamefont {M.}~\bibnamefont
  {Lewenstein}}, \bibinfo {author} {\bibfnamefont {A.}~\bibnamefont {Sanpera}},
  \bibinfo {author} {\bibfnamefont {V.}~\bibnamefont {Ahufinger}}, \bibinfo
  {author} {\bibfnamefont {B.}~\bibnamefont {Damski}}, \bibinfo {author}
  {\bibfnamefont {A.}~\bibnamefont {Sen(De)}},\ and\ \bibinfo {author}
  {\bibfnamefont {U.}~\bibnamefont {Sen}},\ }\bibfield  {title} {\bibinfo
  {title} {Ultracold atomic gases in optical lattices: mimicking condensed
  matter physics and beyond},\ }\href
  {https://doi.org/10.1080/00018730701223200} {\bibfield  {journal} {\bibinfo
  {journal} {Adv. Phys.}\ }\textbf {\bibinfo {volume} {56}},\ \bibinfo {pages}
  {243} (\bibinfo {year} {2007})},\ \Eprint
  {https://arxiv.org/abs/https://doi.org/10.1080/00018730701223200}
  {https://doi.org/10.1080/00018730701223200} \BibitemShut {NoStop}%
\bibitem [{\citenamefont {Louis}\ \emph {et~al.}(2004)\citenamefont {Louis},
  \citenamefont {Ostrovskaya},\ and\ \citenamefont
  {Kivshar}}]{louis:JOptB:2004}%
  \BibitemOpen
  \bibfield  {author} {\bibinfo {author} {\bibfnamefont {P.~J.~Y.}\
  \bibnamefont {Louis}}, \bibinfo {author} {\bibfnamefont {E.~A.}\ \bibnamefont
  {Ostrovskaya}},\ and\ \bibinfo {author} {\bibfnamefont {Y.~S.}\ \bibnamefont
  {Kivshar}},\ }\bibfield  {title} {\bibinfo {title} {Matter-wave dark solitons
  in optical lattices},\ }\href {https://doi.org/10.1088/1464-4266/6/5/020}
  {\bibfield  {journal} {\bibinfo  {journal} {J. Opt. B: Quantum Semiclass.
  Opt.}\ }\textbf {\bibinfo {volume} {6}},\ \bibinfo {pages} {S309} (\bibinfo
  {year} {2004})}\BibitemShut {NoStop}%
\bibitem [{\citenamefont {Grimm}\ \emph {et~al.}(2000)\citenamefont {Grimm},
  \citenamefont {Weidemüller},\ and\ \citenamefont
  {Ovchinnikov}}]{grimm:AdvAtMolOptPhys:2000}%
  \BibitemOpen
  \bibfield  {author} {\bibinfo {author} {\bibfnamefont {R.}~\bibnamefont
  {Grimm}}, \bibinfo {author} {\bibfnamefont {M.}~\bibnamefont
  {Weidemüller}},\ and\ \bibinfo {author} {\bibfnamefont {Y.~B.}\ \bibnamefont
  {Ovchinnikov}},\ }\bibfield  {title} {\bibinfo {title} {Optical dipole traps
  for neutral atoms}\ }(\bibinfo  {publisher} {Academic Press},\ \bibinfo
  {year} {2000})\ pp.\ \bibinfo {pages} {95--170}\BibitemShut {NoStop}%
\bibitem [{\citenamefont {Bloch}\ \emph {et~al.}(2008)\citenamefont {Bloch},
  \citenamefont {Dalibard},\ and\ \citenamefont {Zwerger}}]{bloch:RMP:2008}%
  \BibitemOpen
  \bibfield  {author} {\bibinfo {author} {\bibfnamefont {I.}~\bibnamefont
  {Bloch}}, \bibinfo {author} {\bibfnamefont {J.}~\bibnamefont {Dalibard}},\
  and\ \bibinfo {author} {\bibfnamefont {W.}~\bibnamefont {Zwerger}},\
  }\bibfield  {title} {\bibinfo {title} {Many-body physics with ultracold
  gases},\ }\href {https://doi.org/10.1103/RevModPhys.80.885} {\bibfield
  {journal} {\bibinfo  {journal} {Rev. Mod. Phys.}\ }\textbf {\bibinfo {volume}
  {80}},\ \bibinfo {pages} {885} (\bibinfo {year} {2008})}\BibitemShut
  {NoStop}%
\bibitem [{\citenamefont {Weller}(2009)}]{weller:PhDThesis:2009}%
  \BibitemOpen
  \bibfield  {author} {\bibinfo {author} {\bibfnamefont {A.}~\bibnamefont
  {Weller}},\ }\emph {\bibinfo {title} {Dynamics and Interaction of Dark
  Solitons in Bose-Einstein Condensates}},\ \href@noop {} {Ph.D. thesis},\
  \bibinfo  {school} {University of Heidelberg}, \bibinfo {address} {Germany}
  (\bibinfo {year} {2009})\BibitemShut {NoStop}%
\bibitem [{\citenamefont {Sanz}\ and\ \citenamefont
  {Miret-Art{\'{e}}s}(2008)}]{sanz:JPA:2008}%
  \BibitemOpen
  \bibfield  {author} {\bibinfo {author} {\bibfnamefont {A.~S.}\ \bibnamefont
  {Sanz}}\ and\ \bibinfo {author} {\bibfnamefont {S.}~\bibnamefont
  {Miret-Art{\'{e}}s}},\ }\bibfield  {title} {\bibinfo {title} {A
  trajectory-based understanding of quantum interference},\ }\href
  {https://doi.org/10.1088/1751-8113/41/43/435303} {\bibfield  {journal}
  {\bibinfo  {journal} {J. Phys. A: Math. Theor.}\ }\textbf {\bibinfo {volume}
  {41}},\ \bibinfo {pages} {435303} (\bibinfo {year} {2008})}\BibitemShut
  {NoStop}%
\bibitem [{\citenamefont {Sanz}\ and\ \citenamefont
  {Miret-Art\'es}(2014)}]{sanz-bk-2}%
  \BibitemOpen
  \bibfield  {author} {\bibinfo {author} {\bibfnamefont {A.~S.}\ \bibnamefont
  {Sanz}}\ and\ \bibinfo {author} {\bibfnamefont {S.}~\bibnamefont
  {Miret-Art\'es}},\ }\href@noop {} {\emph {\bibinfo {title} {A Trajectory
  Description of Quantum Processes. II. Applications}}},\ \bibinfo {series}
  {Lecture Notes in Physics}, Vol.\ \bibinfo {volume} {831}\ (\bibinfo
  {publisher} {Springer},\ \bibinfo {address} {Berlin},\ \bibinfo {year}
  {2014})\BibitemShut {NoStop}%
\bibitem [{\citenamefont {Feit}\ and\ \citenamefont
  {J.~A.~Fleck}(1978)}]{feit-fleck:ApplOpt:1978}%
  \BibitemOpen
  \bibfield  {author} {\bibinfo {author} {\bibfnamefont {M.~D.}\ \bibnamefont
  {Feit}}\ and\ \bibinfo {author} {\bibfnamefont {J.}~\bibnamefont
  {J.~A.~Fleck}},\ }\bibfield  {title} {\bibinfo {title} {Light propagation in
  graded-index optical fibers},\ }\href {https://doi.org/10.1364/AO.17.003990}
  {\bibfield  {journal} {\bibinfo  {journal} {Appl. Opt.}\ }\textbf {\bibinfo
  {volume} {17}},\ \bibinfo {pages} {3990} (\bibinfo {year}
  {1978})}\BibitemShut {NoStop}%
\bibitem [{\citenamefont {Feit}\ and\ \citenamefont
  {J.~A.~Fleck}(1980)}]{feit-fleck:ApplOpt:1980-1}%
  \BibitemOpen
  \bibfield  {author} {\bibinfo {author} {\bibfnamefont {M.~D.}\ \bibnamefont
  {Feit}}\ and\ \bibinfo {author} {\bibfnamefont {J.}~\bibnamefont
  {J.~A.~Fleck}},\ }\bibfield  {title} {\bibinfo {title} {Computation of mode
  properties in optical fiber waveguides by a propagating beam method},\ }\href
  {https://doi.org/10.1364/AO.19.001154} {\bibfield  {journal} {\bibinfo
  {journal} {Appl. Opt.}\ }\textbf {\bibinfo {volume} {19}},\ \bibinfo {pages}
  {1154} (\bibinfo {year} {1980})}\BibitemShut {NoStop}%
\bibitem [{\citenamefont {Feit}\ \emph {et~al.}(1982)\citenamefont {Feit},
  \citenamefont {J.~A.~Fleck},\ and\ \citenamefont
  {Steiger}}]{feit-fleck:JCompPhys:1982}%
  \BibitemOpen
  \bibfield  {author} {\bibinfo {author} {\bibfnamefont {M.~D.}\ \bibnamefont
  {Feit}}, \bibinfo {author} {\bibfnamefont {J.}~\bibnamefont {J.~A.~Fleck}},\
  and\ \bibinfo {author} {\bibfnamefont {A.}~\bibnamefont {Steiger}},\
  }\bibfield  {title} {\bibinfo {title} {Solution of the Schr\"odinger equation
  by a spectral method},\ }\href
  {https://doi.org/https://doi.org/10.1016/0021-9991(82)90091-2} {\bibfield
  {journal} {\bibinfo  {journal} {J. Comput. Phys.}\ }\textbf {\bibinfo
  {volume} {47}},\ \bibinfo {pages} {412} (\bibinfo {year} {1982})}\BibitemShut
  {NoStop}%
\bibitem [{\citenamefont {Press}\ \emph {et~al.}(1996)\citenamefont {Press},
  \citenamefont {s.~A.~Teukolsky}, \citenamefont {Vetterling},\ and\
  \citenamefont {Flannery}}]{press-bk-2}%
  \BibitemOpen
  \bibfield  {author} {\bibinfo {author} {\bibfnamefont {W.~H.}\ \bibnamefont
  {Press}}, \bibinfo {author} {\bibnamefont {s.~A.~Teukolsky}}, \bibinfo
  {author} {\bibfnamefont {W.~T.}\ \bibnamefont {Vetterling}},\ and\ \bibinfo
  {author} {\bibfnamefont {B.~P.}\ \bibnamefont {Flannery}},\ }\href@noop {}
  {\emph {\bibinfo {title} {Numerical Recipes in Fortran 90: The Art of
  Parallel Scientific Computing}}},\ \bibinfo {edition} {2nd}\ ed.,\ Vol.\
  \bibinfo {volume} {2nd}\ (\bibinfo  {publisher} {Cambridge University
  Press},\ \bibinfo {address} {Cambridge},\ \bibinfo {year} {1996})\BibitemShut
  {NoStop}%
\bibitem [{\citenamefont {Tonomura}\ \emph {et~al.}(1986)\citenamefont
  {Tonomura}, \citenamefont {Osakabe}, \citenamefont {Matsuda}, \citenamefont
  {Kawasaki}, \citenamefont {Endo}, \citenamefont {Yano},\ and\ \citenamefont
  {Yamada}}]{tonomura:PRL:1986}%
  \BibitemOpen
  \bibfield  {author} {\bibinfo {author} {\bibfnamefont {A.}~\bibnamefont
  {Tonomura}}, \bibinfo {author} {\bibfnamefont {N.}~\bibnamefont {Osakabe}},
  \bibinfo {author} {\bibfnamefont {T.}~\bibnamefont {Matsuda}}, \bibinfo
  {author} {\bibfnamefont {T.}~\bibnamefont {Kawasaki}}, \bibinfo {author}
  {\bibfnamefont {J.}~\bibnamefont {Endo}}, \bibinfo {author} {\bibfnamefont
  {S.}~\bibnamefont {Yano}},\ and\ \bibinfo {author} {\bibfnamefont
  {H.}~\bibnamefont {Yamada}},\ }\bibfield  {title} {\bibinfo {title} {Evidence
  for Aharonov-Bohm effect with magnetic field completely shielded from
  electron wave},\ }\href {https://doi.org/10.1103/PhysRevLett.56.792}
  {\bibfield  {journal} {\bibinfo  {journal} {Phys. Rev. Lett.}\ }\textbf
  {\bibinfo {volume} {56}},\ \bibinfo {pages} {792} (\bibinfo {year}
  {1986})}\BibitemShut {NoStop}%
\bibitem [{\citenamefont {Philippidis}\ \emph {et~al.}(1982)\citenamefont
  {Philippidis}, \citenamefont {Bohm},\ and\ \citenamefont
  {Kaye}}]{philippidis:NuovoCim:1982}%
  \BibitemOpen
  \bibfield  {author} {\bibinfo {author} {\bibfnamefont {C.}~\bibnamefont
  {Philippidis}}, \bibinfo {author} {\bibfnamefont {D.}~\bibnamefont {Bohm}},\
  and\ \bibinfo {author} {\bibfnamefont {R.~D.}\ \bibnamefont {Kaye}},\
  }\bibfield  {title} {\bibinfo {title} {The Aharonov-Bohm effect and the
  quantum potential},\ }\href {https://doi.org/10.1007/BF02721695} {\bibfield
  {journal} {\bibinfo  {journal} {Nuovo Cim. B}\ }\textbf {\bibinfo {volume}
  {71}},\ \bibinfo {pages} {57} (\bibinfo {year} {1982})}\BibitemShut {NoStop}%
\bibitem [{\citenamefont {Sanz}\ and\ \citenamefont
  {Miret-Art\'es}(2007)}]{sanz:cpl:2007}%
  \BibitemOpen
  \bibfield  {author} {\bibinfo {author} {\bibfnamefont {A.~S.}\ \bibnamefont
  {Sanz}}\ and\ \bibinfo {author} {\bibfnamefont {S.}~\bibnamefont
  {Miret-Art\'es}},\ }\bibfield  {title} {\bibinfo {title} {Aspects of
  nonlocality from a quantum trajectory perspective: A WKB approach to Bohmian
  mechanics},\ }\href
  {https://doi.org/https://doi.org/10.1016/j.cplett.2007.08.002} {\bibfield
  {journal} {\bibinfo  {journal} {Chem. Phys. Lett.}\ }\textbf {\bibinfo
  {volume} {445}},\ \bibinfo {pages} {350} (\bibinfo {year}
  {2007})}\BibitemShut {NoStop}%
\end{thebibliography}


%


\newpage

\def\theequation{S\arabic{equation}}
\def\thesection{S\arabic{section}}
\def\thefigure{S\arabic{figure}}
\setcounter{figure}{2}

\begin{widetext}

\section*{Supplemental Material}

\noindent
Figures from the main text without the sets of Bohmian trajectories in order to better
appreciate the corresponding density plots.
\vspace{2cm}
\begin{figure*}[h]
 \centering
 \includegraphics[width=\textwidth]{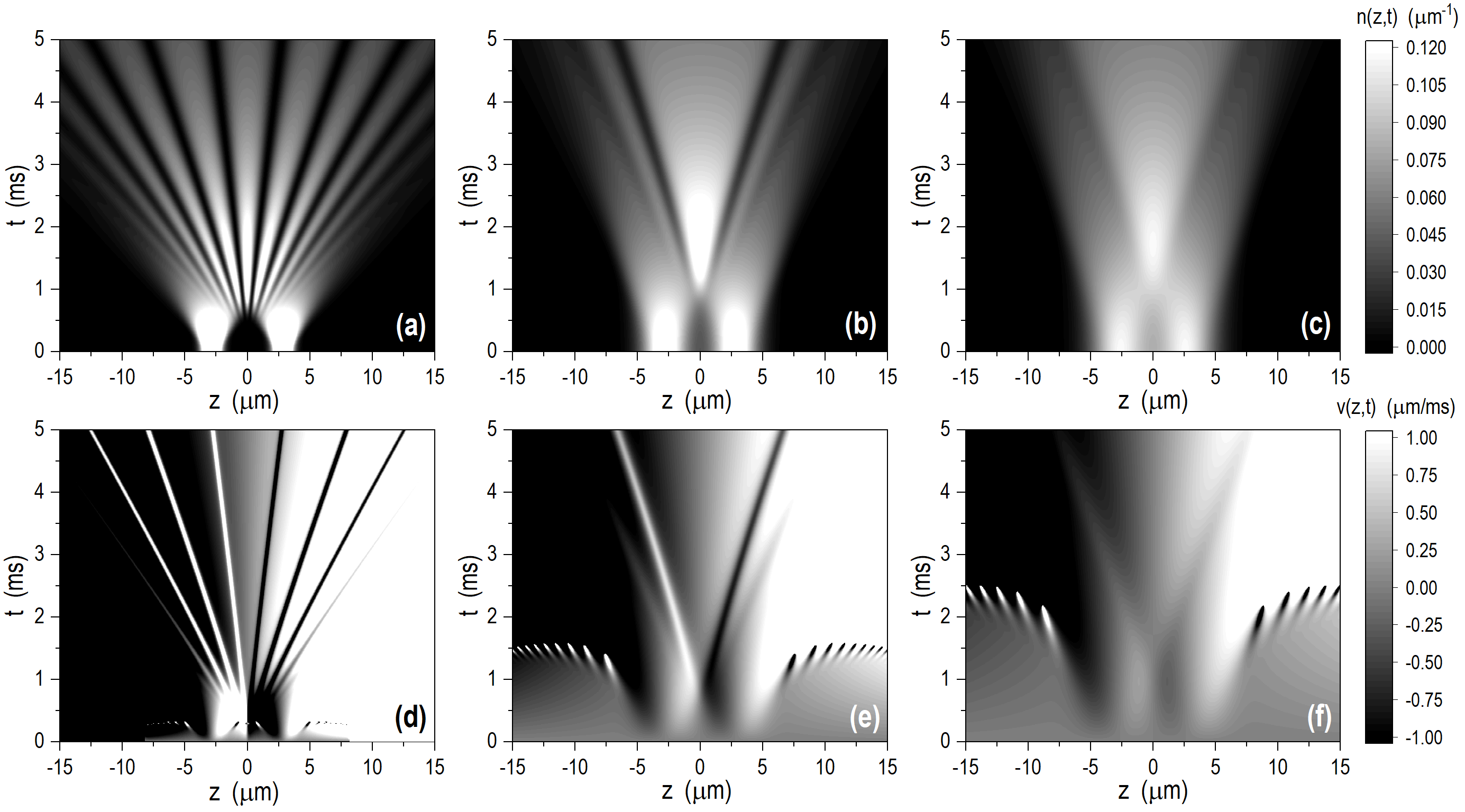}
 \caption{\label{figS3}
  Time-evolution of the BEC density distribution (top row) and its associated velocity field
  (lower row) for the three cases displayed in Fig.~2: (a/d) $r = 1$, (b/e) $r = 2.5$,
  and (c/f) $r = 3.2$.
  The color code to the right denotes the scale considered, from black for the lowest values of the
  quantities displayed to white for the highest values considered.
  For a better visualization and comparison, the density distributions have been truncated to
  0.12~$\mu$m$^{-3}$, while the velocity field ranges within $\pm 1$~$\mu$m/ms.}
\end{figure*}

\newpage

\phantom{text}
\vspace{4cm}
\begin{figure*}[h]
\centering
 \includegraphics[width=\textwidth]{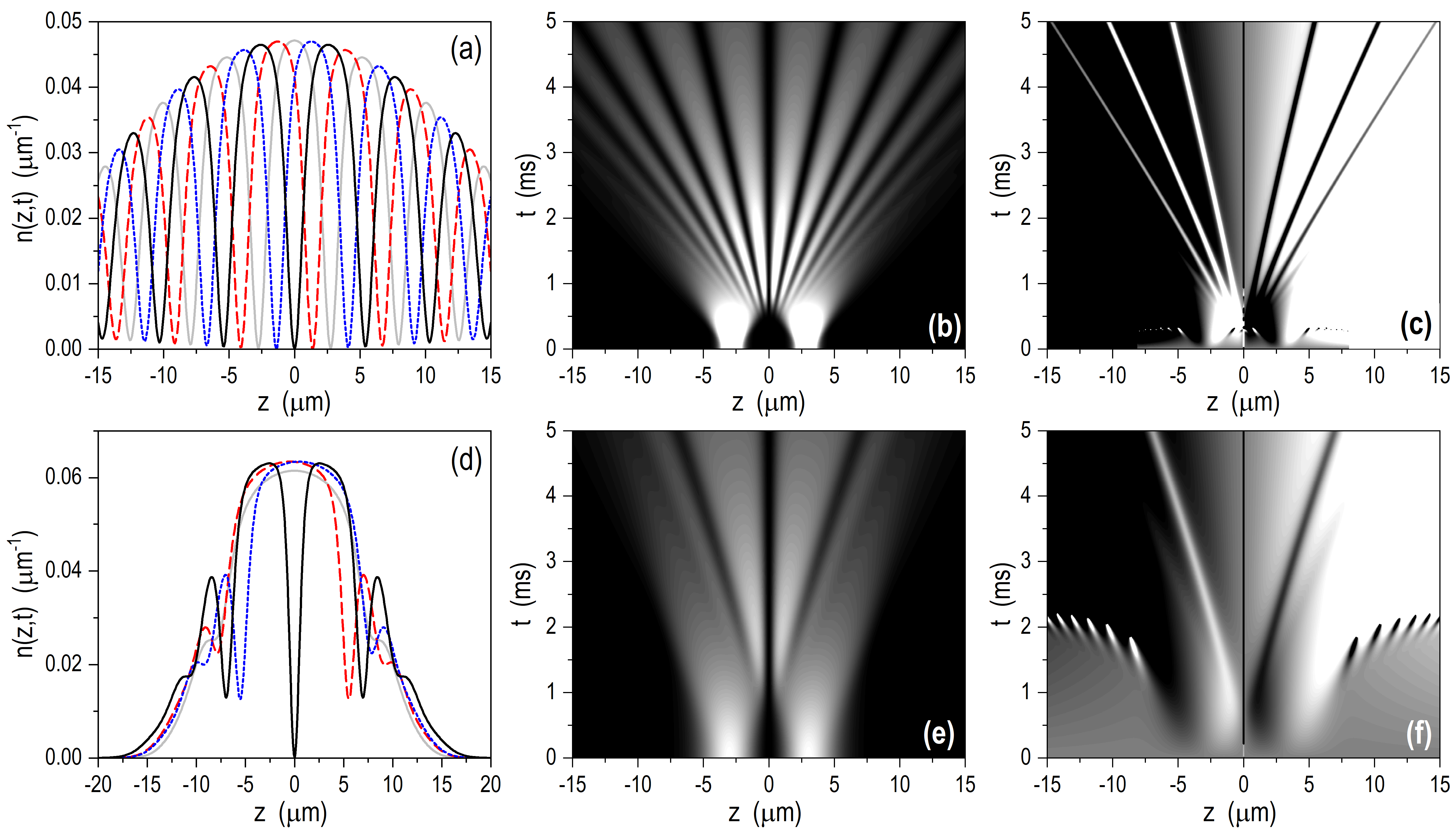}
 \caption{\label{figS4}
  Effect of the phase difference $\phi$ on the dynamics exhibited by a coherent
  superposition of two condensates with identical Gaussian distributions, with peak-to-peak
  distance $\ell = 5.7$~$\mu$m and width $\sigma_0 = r \sigma_{\rm eff} \approx 0.49 r$~$\mu$m.
  In the upper row, results for $r = 1$: (a) density profiles at $t = 5$~ms for $\phi = 0$
  (solid gray line), $\pi/2$ (dashed red line), $\pi$ (solid black line), and $-\pi/2$
  (dotted blue line); (b) and (c) density plots showing the time evolution of the density
  distribution and the velocity field, respectively.
  In the lower row, the same for $r=3.2$.
  The color code for panels (b), (c), (d), and (f) is the same as in Fig.~S3.}
\end{figure*}

\newpage

\phantom{text}
\vspace{3cm}
\begin{figure*}[h]
\centering
 \includegraphics[width=\textwidth]{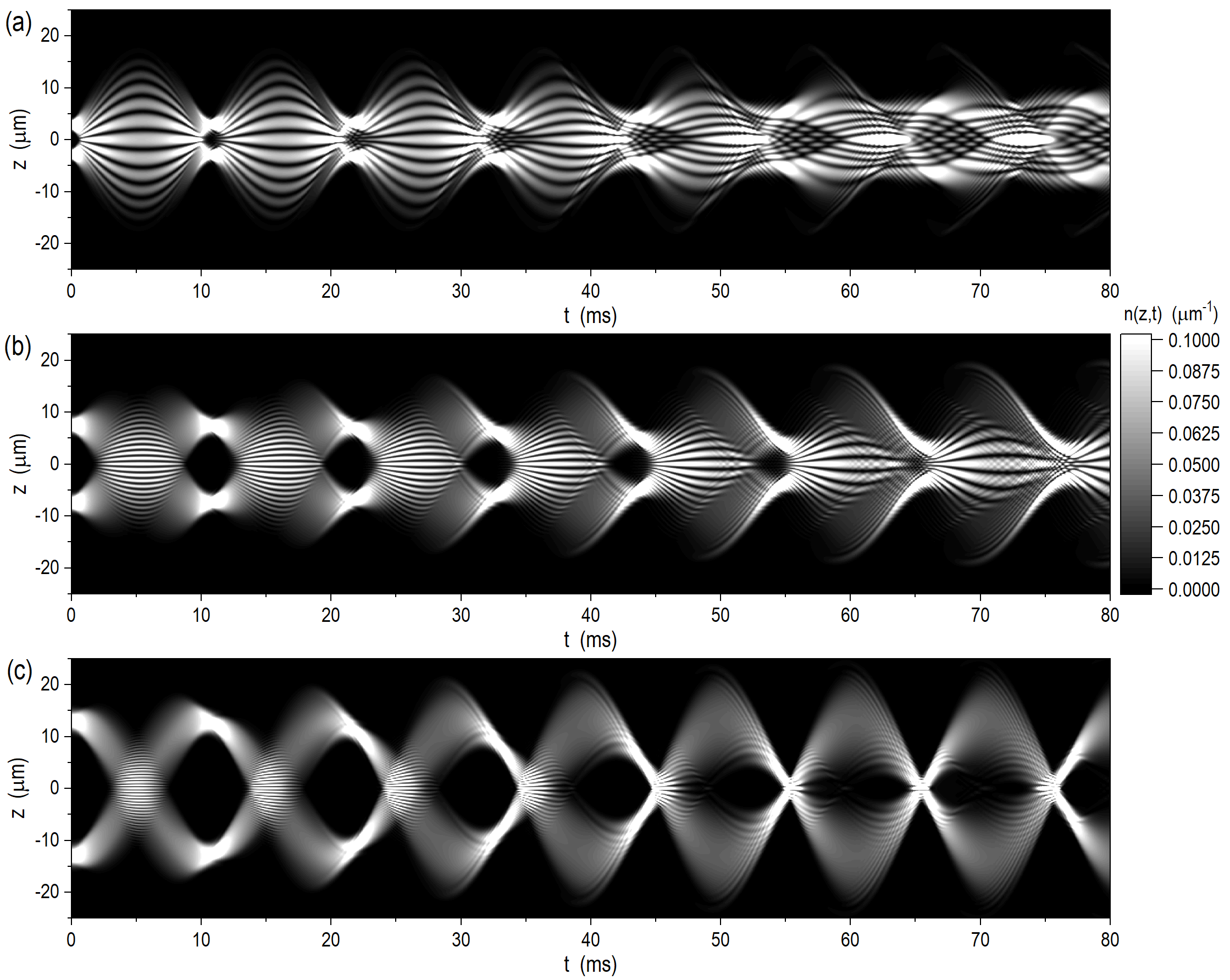}
 \caption{\label{figS5}
  Contour plots showing the time evolution of the density distribution associated
  with three initial superpositions with peak-to-peak distances:
  (a) $\ell = 5.7$~$\mu$m, (b) $\ell = 15$~$\mu$m, and (c) $\ell = 26$~$\mu$m.
  The color code is defined on the right; the distributions have been truncated
  to $n_{\rm max} = 0.1$~$\mu$m$^{-1}$ in all cases for a better visualization.
  In all cases the frequency of the harmonic trap is $f_z = 50$~Hz.}
\end{figure*}

\newpage

\phantom{text}
\vspace{3cm}
\begin{figure*}[h]
 \centering
 \includegraphics[width=\textwidth]{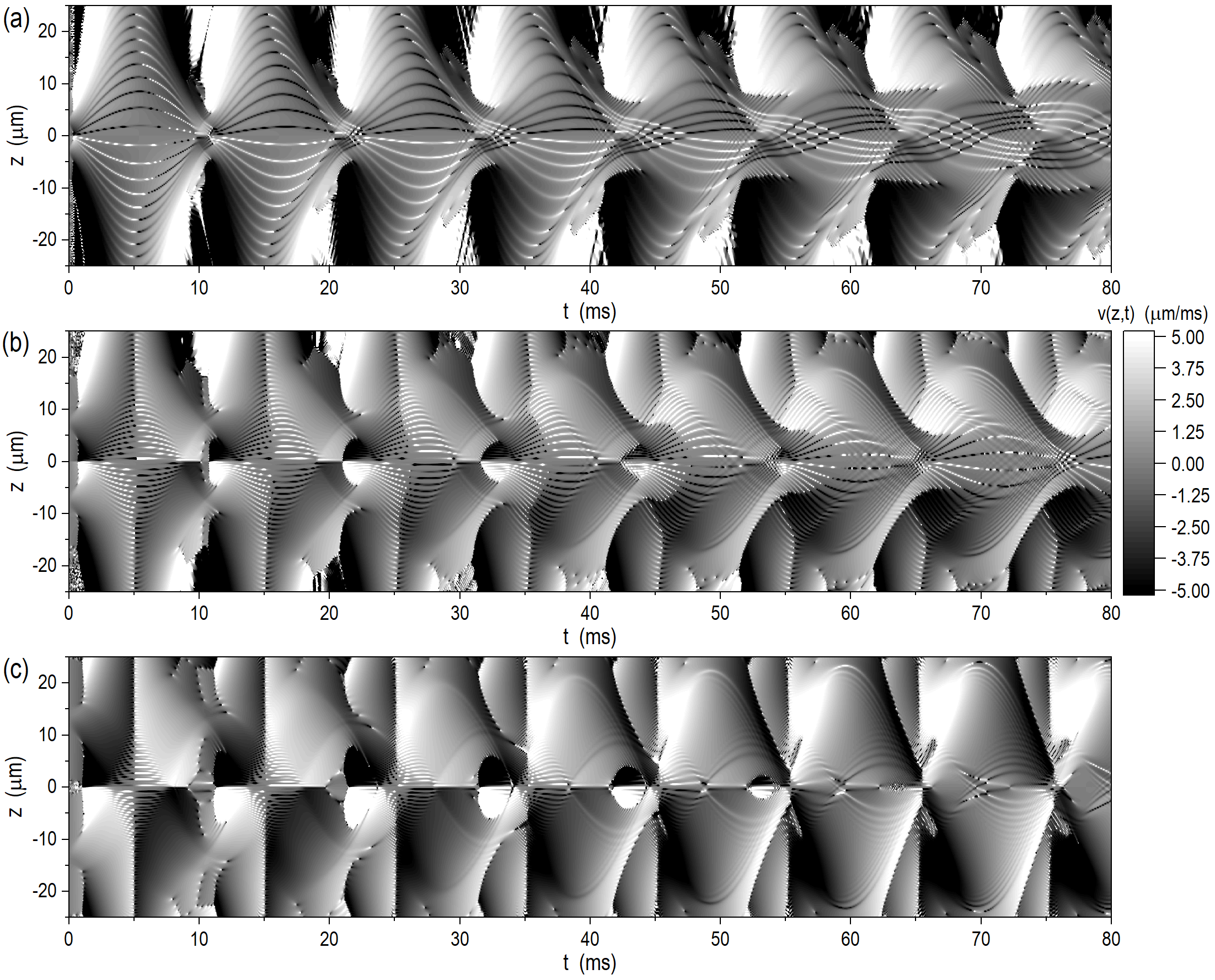}
 \caption{\label{figS6}
  Contour plots showing the time evolution of the velocity field associated with
  the three initial superpositions considered in Fig.~5, with
  peak-to-peak distances: (a) $\ell = 5.7$~$\mu$m, (b) $\ell = 15$~$\mu$m,
  and (c) $\ell = 26$~$\mu$m.
  The color code is defined on the right; velocity values have been constrained
  to the range $\pm 5$~$\mu$m/ms in all cases for a better visualization.
  In all cases the frequency of the harmonic trap is $f_z = 50$~Hz.}
\end{figure*}

\end{widetext}

\end{document}